%% file: DegradingUpgrading_arXiv.tex
\pdfoutput=1
\documentclass[final,journal, twoside, 10pt]{IEEEtran}
\usepackage[hyperref]{Bamble}
\input{DegradingUpgrading_Shortcuts}

\begin{document}

\title{Greedy-Merge Degrading has Optimal Power-Law}
\author{Assaf~Kartowsky and~Ido~Tal\\  
       Department of Electrical Engineering \\
Technion - Haifa 32000, Israel\\
E-mail: \{kartov@campus, idotal@ee\}.technion.ac.il
}
\maketitle

\begin{abstract}
Consider a channel with a given input alphabet size and a given input distribution.
Our aim is to degrade or upgrade it to a channel with at most $L$ output letters.

The paper contains four main results. The first result, from which the paper title is derived, deals with the so called ``greedy-merge'' algorithm. 
We derive an upper bound on the reduction in mutual information between input and output.
This upper bound is within a constant factor of an algorithm-independent lower bound.
Thus, we establish that greedy-merge is optimal in the power-law sense.

The other main results deal with upgrading. The second result shows that a certain sequence of channels that was previously shown to be ``hard'' for degrading, displays the same hardness in the context of upgrading. 
That is, suppose we are given such a channel and a corresponding input distribution. 
If we upgrade (degrade) to a new channel with $L$ output letters, we incur an increase (decrease) in mutual information between input and output. 
We show that a previously derived bound on the decrease in mutual information for the degrading case is also a lower bound on the increase for the upgrading case.

The third result is an efficient algorithm for optimal upgrading, in the binary-input case. 
That is, we are given a channel and an input distribution. 
We must find an upgraded channel with $L$ output letters, for which the increase in mutual information is minimal. 
We give a simple characterization of such a channel, which implies an efficient algorithm.

The fourth result is an analog of the first result for the upgrading case, when the input is binary. 
That is, we first present a sub-optimal algorithm for the setting considered in the third result. 
The main advantage of the sub-optimal algorithm is that it is amenable to analysis. 
We carry out the analysis, and show that the increase incurred in mutual information is within a constant factor of the lower bound derived in the second result. 

\end{abstract}
\section{Introduction}
\label{sec:Introduction}
\IEEEPARstart{I}{n} myriad digital processing contexts, quantization is used to map a large alphabet to a smaller one. 
For example, quantizers are an essential building block in receiver design, used to keep the complexity and resource consumption manageable. 
The quantizer used has a direct influence on the attainable code rate. 

Another recent application is related to polar codes~\cite{Arikan:PolarCodes}. 
Polar code construction is equivalent to evaluating the misdecoding probability of each channel in a set of synthetic channels. 
This evaluation cannot be carried out naively, since the output alphabet size of a synthetic channel is intractably large. 
One approach to circumvent this difficulty is to degrade and upgrade the evaluated synthetic channel to a channel with manageable output alphabet size \cite{Tal:Construct}. 
Thus, one obtains upper and lower bounds on the misdecoding probability. 
In particular, for a wiretap channel, both upgrading and degrading are essential to ensure secure and reliable communications~\cite{MahdavifarVardy:11p}\cite{HofShamai:10a}\cite{ARTKS:10p}\cite{KoyluogluElGamal:10c}\cite{Tal:Wiretap}.

The general problem considered in this paper is the following. 
Given a design parameter $L$, we either degrade or upgrade an initial channel to a new one with output alphabet size at most $L$. 
We assume that the input distribution is specified, and note that degradation reduces the mutual information between the channel input and output, whereas upgradation increases it. 
This reduction (increase) in mutual information is roughly the loss (gain) in code rate due to quantization. We denote the smallest reduction (increase) possible by $\DIDegStar$ ($\DIUpgStar$) or simply $\Delta I^\ast$, when direction is clear from the context.

In this work we present four main results. For the sake of clarity, \Cref{tbl:PriorWorks} lists previous related works along with our new results, marked by a ``$\checkmark$''. In the table, BDMC and DMC are short for Binary Discrete Memoryless Channel and Discrete Memoryless Channel, respectively. We note that $\Delta I^\ast=\Omega(\cdot)$ denotes lower bounds on $\Delta I^\ast$ as a function of $L$ for a specified channel and input distribution or a sequence of those two. On the other hand, $\Delta I^\ast=O(\cdot)$ are general upper bounds on $\Delta I^\ast$, as a function of $L$, that are independent of channel and input distribution.
\begin{table}[h]
	\caption{Previous related works and our new results}
	\label{tbl:PriorWorks}	
	\centering
	\begin{tabular}{ccccc}
		\hline
									&	Channel	&	Optimal							&	\multirow{2}{*}{$\Delta I^\ast=\Omega(\cdot)$}	&	\multirow{2}{*}{$\Delta I^\ast=O(\cdot)$}\\
									&	Type	&	Algorithm						&													&	 \\
		\hline
		\multirow{2}{*}{Degrading}	&	BDMC	&	\cite{Kurkoski:Quantization},\cite{Iwata:SMAWK}	&	\cite{Tal:ModerateSizes}						&	\cite{Tal:Construct},\cite{Pedarsani:Construction}\footnotemark\addtocounter{footnote}{-1}\addtocounter{Hfootnote}{-1},\checkmark \\ \cline{2-5}
									&	DMC		&		&	\cite{Tal:ModerateSizes}						&	\cite{Tal:ConstructingPolarCodes},\cite{Gulcu:DMC},\cite{PeregTal:Upgrading},\checkmark \\ 
		\hline
		\multirow{2}{*}{Upgrading}	&	BDMC	&	\checkmark						&	\checkmark										&	\cite{Tal:Construct},\cite{Pedarsani:Construction}\footnotemark,\checkmark  \\ \cline{2-5}
									&	DMC		&									&	\checkmark										&	\cite{PeregTal:Upgrading} \\ \hline	
	\end{tabular}
\end{table}
\footnotetext{To be precise, the results in~\cite{Tal:Construct} and~\cite{Pedarsani:Construction} were derived for BMSCs.}

Let $|\calX|$ denote the channel input alphabet size, and treat it as a fixed quantity. 
In our first main result (\Cref{sec:UBDC}), we show that for any input distribution and any initial channel, $\DIDegStar=O(L^{-2/(|\calX|-1)})$. 
Moreover, this bound is attained efficiently by the greedy-merge algorithm discussed already in~\cite{Tal:Construct} and~\cite{Pedarsani:Construction} for binary-input memoryless symmetric channels (BMSCs), and in~\cite{Gulcu:DMC} for general discrete memoryless channels. 
This bound is tighter than the bounds derived in \cite{Tal:Construct},\cite{Pedarsani:Construction},\cite{Tal:ConstructingPolarCodes},\cite{Gulcu:DMC} and \cite{PeregTal:Upgrading}. 
In fact, up to a constant multiplier (dependent on $|\calX|$), this bound is the tightest possible. Namely, \cite{Tal:ModerateSizes} proves the existence of an input distribution and a sequence of channels for which $\DIDegStar=\Omega(L^{-2/(|\calX|-1)})$. 
Both bounds have $-2/(|\calX|-1)$ as the power of $L$, the same power-law. 
We mention a recent result \cite{Nazer:Distilling} in which a different power-law is shown to be tight, for the special case in which the channel is very noisy.

Our second main result (\Cref{sec:LBUC}) is the analog of~\cite{Tal:ModerateSizes} to the upgrading problem. 
Namely, in \cite{Tal:ModerateSizes}, a sequence of channels is shown to have a degrading penalty of $\DIDegStar=\Omega(L^{-2/(|\calX|-1)})$. 
We show that this same sequence of channels has an upgrading penalty $\DIUpgStar=\Omega(L^{-2/(|\calX|-1)})$, with the exact same constant. 
Similar to~\cite{Tal:ModerateSizes}, we conclude that some channels with moderate $|\calX|$ are ``hard'' to upgrade, in the sense that a very large $L$ is required to keep $\DIUpgStar$ small. 
Moreover, this result plays an important role in our fourth result.

An optimal degrading algorithm was presented in \cite{Kurkoski:Quantization}, for the binary-input case. Namely, a channel and input distribution are supplied, along with a target output alphabet size. 
The algorithm finds a degraded channel with the target output alphabet size, that has the largest possible mutual information between input and output. 
See also \cite{Iwata:SMAWK} for an improved implementation in terms of running time. In our third result (\Cref{sec:OBU}), we present the upgrading analog: an optimal upgrading algorithm for binary-input discrete memoryless channels. 
Namely, we show that an optimal upgraded channel is a subset of the initial channel, when both are represented using posterior probability vectors. 
This characterization paves the way for an algorithm that efficiently finds the optimal subset.

In our fourth main result (\Cref{sec:UBBUC}), we use our previous results and techniques to obtain an upper bound on $\DIUpgStar$, valid for any binary-input discrete memoryless channel and any input distribution. 
That is, a greedy version of the optimal upgrading algorithm, known as ``greedy-split'',  is proved to obtain $\DIUpgStar=O(L^{-2/(|\calX|-1)})=O(L^{-2})$. 
We note that the algorithm is a generalization of the one presented in \cite{Tal:Construct} for the symmetric case. 
Our bound is tighter than the one previously derived in~\cite{Pedarsani:Construction}. 
As in our first result, this new bound shares the same power-law as the lower bound from our second result, and is thus, up to a constant multiplier, the tightest possible.

\section{Framework and notation}
We are given an input distribution and a DMC $W: \calX \rightarrow \calY$. 
Both $|\calX|$ and $|\calY|$ are assumed finite. 
Let $X$ and $Y$ denote the random variables that correspond to the channel input and output, respectively. 
Denote the corresponding distributions $P_X$ and $P_Y$. 
The probability of receiving $y \in \calY$ as the channel output given that $x \in \calX$ was transmitted, namely $\Prob{Y=y|X=x}$, is denoted by $W(y|x)$. 
The probability that $x \in \calX$ is the channel input, namely $\Prob{X=x}$, is denoted by $\pi (x)$. 
We also assume that $\calX$ and $\calY$ are disjoint, allowing ourselves to abuse notation and denote $\Prob{X=x|Y=y}$ and $\Prob{Y=y}$ as $W(x|y)$ and $\pi(y)$ respectively. 
We stress that unless stated otherwise, we will not assume that $W$ is symmetric in any sense. 
Without loss of generality we assume that $\pi(x) > 0$ and $\pi(y) > 0$ for every $x \in \calX$ and $y \in \calY$.

The mutual information between the channel input and output is
\[
I(W) \triangleq I(X;Y) = \sum_{x\in \calX} \eta(\pi (x))-\sum_{\substack{x \in \calX, \\ y \in \calY} } \pi(y) \eta(W(x|y)) \; ,
\]
where
\begin{equation}
\label{eq:etaDef}
\eta(p) = \begin{cases}
-p \log p & p > 0 \; , \\
0 & p = 0 \; ,
\end{cases}
\end{equation}
and the logarithm is taken in the natural basis. 
We note that the input distribution does not necessarily have to be the one that achieves the channel capacity.

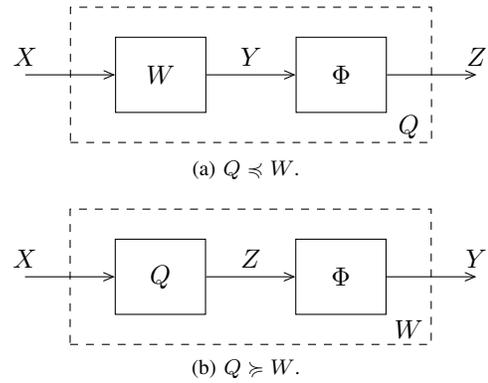
\begin{figure}[t]
	\centering
	\subfloat[$Q \preccurlyeq W$. \label{fig:Degrading}]{
		\begin{tikzpicture}[>=angle 45]
		\node (W) at (-1.2,0) [rectangle, draw ,minimum width = 1.2cm, minimum height = 1cm] {$W$}; 
		\node (P) at (1.2,0) [rectangle, draw, minimum width = 1.2cm, minimum height = 1cm] {$\Phi$}; 
		\draw[->] (W) -- node [above] {$Y$} (P); 
		\draw[<-] (W) -- ++ (-1.8,0) node[above] (u) {$X$}; 
		\draw[->] (P) -- ++ (1.8,0)  node[above] (y) {$Z$}; 
		\node (C) at (0,0) [rectangle, black, draw, minimum width = 4.8 cm, minimum height = 1.8 cm, dashed] {};
		\node(Q) at (2.1,-0.7){$Q$};
		\end{tikzpicture} 
	} \\
	\subfloat[$Q \succcurlyeq W$. \label{fig:Upgrading}]{
		\begin{tikzpicture}[>=angle 45]
		\node (Q) at (-1.2,0) [rectangle, draw ,minimum width = 1.2cm, minimum height = 1cm] {$Q$}; 
		\node (P) at (1.2,0) [rectangle, draw, minimum width = 1.2cm, minimum height = 1cm] {$\Phi$}; 
		\draw[->] (Q) -- node [above] {$Z$} (P); 
		\draw[<-] (Q) -- ++ (-1.8,0) node[above] (u) {$X$}; 
		\draw[->] (P) -- ++ (1.8,0)  node[above] (y) {$Y$}; 
		\node (C) at (0,0) [rectangle, black, draw, minimum width = 4.8 cm, minimum height = 1.8 cm, dashed] {};
		\node(W) at (2.1,-0.7){$W$};
		\end{tikzpicture} 
	}
	\caption{Degrading and upgrading $W$ to $Q$.}
\end{figure}

We now define the relations of degradedness and upgradedness between channels. 
A channel $Q: \calX \rightarrow \calZ$ is said to be \emph{degraded} with respect to a channel $W: \calX \rightarrow \calY$, and we write $Q \preccurlyeq W$, if there exists a channel $\Phi: \calY \rightarrow \calZ$ such that
\begin{equation}
\label{eq:Degrading}
Q(z|x)=\sum_{y \in \calY} W(y|x) \Phi(z|y)
\end{equation}
for all $x \in \calX$ and $z \in \calZ$ (see~\Cref{fig:Degrading}). 
The channel $\Phi$ is then defined as the \emph{intermediate channel}. 
Conversely, the channel $Q$ is said to be \emph{upgraded} with respect to $W$, if $W$ is degraded with respect to $Q$ (see~\Cref{fig:Upgrading}). 
We note that as a result of the data processing theorem, $Q \preccurlyeq W$ implies $\DIDeg \triangleq I(W)-I(Q) \geq 0$, and similarly, $W \preccurlyeq Q$ implies $\DIUpg \triangleq I(Q)-I(W) \geq 0$.

Since our focus is on approximating channels with a large output alphabet size using channels with a limited output alphabet size we define the \emph{optimal degrading loss} for a given pair $(W,P_X)$ and a target output alphabet size $L$ as
\begin{equation}
\label{eq:DIDeg}
\DIDegStar(W,P_X, L)  = \DIDegStar \triangleq \min_{\substack{{Q,\Phi:Q \preccurlyeq W,} \\ {|Q| \leq L}} } I(W)-I(Q) \; ,
\end{equation}
where $|Q|$ denotes the output alphabet size of the channel $Q$. 
The optimizer $Q$ is the degraded channel that is ``closest" to $W$ in the sense of mutual information, yet has at most $L$ output letters. 
In the same manner, we define the \emph{optimal upgrading gain} for a given pair $(W,P_X)$ and a target output alphabet size $L$ as
\begin{equation}
\label{eq:DIUpgStar}
\DIUpgStar(W,P_X, L)  = \DIUpgStar  \triangleq \min_{\substack{{Q,\Phi:Q \succcurlyeq W,} \\ {|Q| \leq L}} } I(Q)-I(W) \; .
\end{equation}
As in~\cite{Tal:ModerateSizes}, define the \emph{degrading cost} in the following way:
\[
\DC(|\calX|,L) \triangleq \sup_{W,P_X} \DIDegStar \; .
\]
The optimizers $W$ and $P_X$ are the channel and input distribution that yield the highest optimal degrading loss. 
In a way, they are the ``worst" or ``hardest" pair to degrade. 
We define the similar notion for the case of upgrading, namely the \emph{upgrading cost} as
\begin{equation}
\label{eq:UCDef}
\UC(|\calX|,L) \triangleq \sup_{W,P_X} \DIUpgStar \; .
\end{equation}

\section{Upper bound on optimal degrading loss}
\label{sec:UBDC}
\subsection{Main result}
Our first main result is an upper bound on $\DIDegStar$ and $\DC$ in terms of $|\calX|$ and $L$ that is tight in the power-law sense. 
This upper bound will follow from analyzing a sub-optimal degrading algorithm, called ``greedy-merge''. 
In each iteration of greedy-merge,  we merge the two output letters $\alpha,\beta \in \calY$ that results in the smallest decrease of mutual information between input and output, denoted $\DIDeg$. 
Namely, the intermediate channel $\Phi$ maps $\alpha$ and $\beta$ to a new symbol, while all other symbols are unchanged by $\Phi$. 
This is repeated $|\calY|-L$ times, to yield an output alphabet size of $L$. 
By upper bounding the $\DIDeg$ of each iteration, we obtain an upper bound on $\DIDegStar$ and $\DC$. 
\begin{theorem}
	\label{thm:DIDegStar}
	Let a DMC $W:\calX \rightarrow \calY$ satisfy $|\calY|>2|\calX|$ and let $L \geq 2|\calX|$. Then, for any fixed input distribution,
	\[
	\DIDegStar =\min_{\substack{{Q,\Phi:Q \preccurlyeq W,} \\ {|Q| \leq L}} } I(W)-I(Q)= O\p{L^{-\frac{2}{|\calX|-1}}} \; .
	\] 
	In particular, 
	\[
	\DIDegStar \leq \nu(|\calX|)\cdot L^{-\frac{2}{|\calX|-1}} \; , 
	\]
	where 
	\begin{multline*}
	\nu(|\calX|) \triangleq \frac{\pi|\calX|(|\calX|-1)}{2\p{\sqrt{1+\frac{1}{2(|\calX|-1)}}-1}^2}  \\  
	\cdot \p{\frac{2|\calX|}{\Gamma \p{1+\frac{|\calX|-1}{2}}}}^{\frac{2}{|\calX|-1}} \; ,
	\end{multline*}
	and $\Gamma(\cdot)$ is the Gamma function. That is, for an integer $n\geq 1$,
	\begin{equation}
	\label{eq:Gamma}
	\Gamma (n)=(n-1)!\;, \quad \Gamma \left( n+\frac{1}{2} \right)=\frac{(2n)!}{4^n n!}\sqrt{\pi}\; .
	\end{equation}
	This bound is attained by greedy-merge, and is tight in the power-law sense.
\end{theorem}
Note that for large values of $|\calX|$ the Stirling approximation along with some other first order approximations can be applied to simplify $\nu(|\calX|)$ to
\[
\nu(|\calX|) \approx  16\pi e |\calX|^3 \; .
\]

We stress that the greedy-merge algorithm is not optimal in the sense of $\DIDegStar$.
To see this, consider the following example.
\begin{example}
	Let $\calX=\{0,1\}$, $\calY=\{a,b,c,d\}$, and let the channel $W:\calX \to \calY$ satisfy
	\begin{align*}
		W(a|0)=0,\;\quad  &W(b|0)=\frac{1}{6},\quad W(c|0)=\frac{1}{3},\quad  W(d|0)= \frac{1}{2}, \\
		W(a|1)=\frac{1}{2},\quad &W(b|1) = \frac{1}{3},\quad W(c|1)=\frac{1}{6},\quad W(d|1) = 0.
	\end{align*}
	Assume also that $\pi(x)=\frac{1}{2}$ for all $x \in \calX$.
	Suppose now that we wish to reduce the output alphabet from $4$ to $2$ (i.e., $L=2$) by degrading. 
	The greedy-merge algorithm would first merge the pair $b,c$ to a new letter $bc$, and then merge $a$ with $bc$, resulting in $\DIDeg=0.16$ in total.
	The optimal degrading algorithm, however, would merge $a,b$ to $ab$ and $c,d$ to $cd$, resulting in a total reduction of $\DIDegStar=0.13$.
	Clearly, the greedy-merge algorithm is not optimal.
\end{example}

To prove \Cref{thm:DIDegStar}, we will use the following theorem which proves the existence of a pair of output letters whose merger yields a ``small'' $\DIDeg$.
\begin{theorem}
\label{thm:DIDeg}
	Let a DMC $W:\calX \rightarrow \calY$ satisfy $|\calY|>2|\calX|$, and let the input distribution be fixed. 
	There exists a pair $\alpha,\beta \in \calY$ whose merger results in a channel $Q$ satisfying 
	\[
	\DIDeg = O\p{|\calY|^{-\frac{|\calX|+1}{|\calX|-1}}} \; .
	\]
	In particular,
	\begin{equation}
	\label{eq:DIDegOrder}
	\DIDeg \leq   \mu(|\calX|)\cdot|\calY|^{-\frac{|\calX|+1}{|\calX|-1}} \; ,
	\end{equation}
	where,
	\[
	\mu(|\calX|) \triangleq \frac{2}{|\calX|-1} \nu(|\calX|) \; ,
	\]
	and $\nu(\cdot)$ was defined in \Cref{thm:DIDegStar}.
\end{theorem}

We first prove \Cref{thm:DIDeg} throughout the section and then prove \Cref{thm:DIDegStar}.

\subsection{An alternative ``distance" function}
We begin by addressing the merger of a pair of output letters $\alpha,\beta \in \calY$. 
Since this pair of letters is merged into a new letter $\gamma$ (we assume $\gamma \notin \calX \cup \calY$), the new output alphabet of $Q$ is $\calZ=\calY \setminus \ppp{\alpha,\beta} \cup \ppp{\gamma}$. 
The channel $Q:\calX \rightarrow \calZ$ then satisfies
\[
Q(\gamma|x)=W(\alpha|x)+W(\beta|x) \; ,
\]
whereas for all $y \in \calZ \cap \calY$ we have $Q(y|x)=W(y|x)$. 
Using the shorthand
\[
\pi_{\gamma} = \pi(\gamma) \; , \quad \pi_{\alpha} = \pi(\alpha) \; , \quad \pi_{\beta} = \pi(\beta) \; ,
\]
one gets that
\[
\pi_{\gamma}=\pi_{\alpha}+\pi_{\beta} \; .
\]
Let us denote by $\bfalpha=(\alpha_x)_{x \in \calX}$, $\bfbeta=(\beta_x)_{x \in \calX}$ and $\bfgamma=(\gamma_x)_{x \in \calX}$ the vectors corresponding to posterior probabilities associated with $\alpha,\beta$ and $\gamma$, respectively. 
Namely,
\[
\alpha_x = W(x|\alpha) \; , \quad \beta_x = W(x|\beta) \; ,
\]
and a short calculation shows that
\begin{equation}
\label{eq:gamma}
\gamma_x = Q(x|\gamma) = \frac{\pi_{\alpha} \alpha_x + \pi_{\beta} \beta_x }{\pi_{\gamma}}=\frac{\pi_{\alpha} \alpha_x + \pi_{\beta} \beta_x }{\pi_{\alpha}+\pi_{\beta}} \; . 
\end{equation}
Thus, after canceling terms, one gets that
\begin{equation}
\label{eq:DIxsum}
\DIDeg=I(W)-I(Q)=\sum_{x \in \calX} \Delta I_x \; ,
\end{equation}
where
\[
\Delta I_x \triangleq \pi_{\gamma}\eta(\gamma_x)-\pi_{\alpha}\eta(\alpha_x)-\pi_{\beta}\eta(\beta_x) \; .
\]

In order to bound $\DIDeg$, we give two bounds on $\Delta I_x$. 
The first bound was derived in~\cite{Gulcu:DMC},
\begin{equation}
\label{eq:d1}
	\Delta I_x \leq (\pi_{\alpha} + \pi_{\beta})\cdot d_1(\alpha_x,\beta_x) \; ,
\end{equation}
where for $\nonnegvar \geq 0$ and $\zeta \in \reals$, we define
\[
d_1 (\nonnegvar,\zeta) \triangleq |\zeta-\nonnegvar| \; .
\]

The subscript ``$1$'' in $d_1$ is suggestive of the $L_1$ distance. 
Note that we will generally use $\alpha_x$ or $\nonnegvar$ to denote a probability associated with an input letter, while $\zeta$ will denote a ``free'' real variable, possibly negative. 
We will keep to this convention for the vector case as well. In addition, let us point out that the bound in \eqref{eq:d1} was derived assuming a uniform input distribution, however it remains valid for the general case.

We now derive the second bound on $\Delta I_x$. 
For the case where $\alpha_x,\beta_x >0$,
\begin{align*}
	\Delta I_x &= \pi_{\alpha} (\eta(\gamma_x)-\eta(\alpha_x)) + \pi_{\beta}(\eta(\gamma_x)-\eta(\beta_x)) \\
	&\stackrel{(a)}{\leq} \pi_{\alpha} \eta'(\alpha_x)(\gamma_x-\alpha_x) + \pi_{\beta} \eta'(\beta_x)(\gamma_x-\beta_x) \\
	&\stackrel{(b)}{=} \frac{\pi_{\alpha}\pi_{\beta}}{\pi_{\alpha}+\pi_{\beta}}(\alpha_x-\beta_x)(\eta'(\beta_x)-\eta'(\alpha_x))\\
	&\stackrel{(c)}{\leq} \frac{1}{4}(\pi_{\alpha}+\pi_{\beta})(\alpha_x-\beta_x)^2 (-\eta''(\lambda)) \; ,
\end{align*}
where in $(a)$ we used the concavity of $\eta(\cdot)$, in $(b)$ the definition of $\gamma_x$ (see \eqref{eq:gamma}), and in $(c)$ the AM-GM inequality and the mean value theorem where $\lambda=\theta\alpha_x +(1-\theta)\beta_x$ for some $\theta \in [0,1]$. Using the monotonicity of $-\eta''(p)=1/p$ we get
\[
-\eta''(\lambda) \leq \frac{1}{\min(\alpha_x,\beta_x)} \; .
\]
Thus,
\begin{equation}
\label{eq:d2}
	\Delta I_x \leq (\pi_{\alpha}+\pi_{\beta}) \cdot d_2(\alpha_x,\beta_x) \; ,
\end{equation}
where 
\[
d_2(\nonnegvar,\zeta) \triangleq \begin{cases}
\frac{(\zeta-\nonnegvar)^2}{\min(\nonnegvar,\zeta)} & \nonnegvar,\zeta > 0 \; , \\
\infty & \mbox{otherwise} \; .
\end{cases}
\]
The subscript ``$2$'' in $d_2$ is suggestive of the squaring in the numerator. 
Combining \eqref{eq:d1} and \eqref{eq:d2} yields
\begin{equation}
\label{eq:d}
	\Delta I_x \leq (\pi_{\alpha}+\pi_{\beta}) \cdot d(\alpha_x,\beta_x) \; ,
\end{equation}
where
\begin{equation}
\label{eq:dScalarDef}
d(\nonnegvar,\zeta) \triangleq \min(d_1(\nonnegvar,\zeta),d_2(\nonnegvar,\zeta)) \; .
\end{equation}
Returning to \eqref{eq:DIxsum} using \eqref{eq:d} we get
\begin{equation}
\label{eq:DIBound}
	\DIDeg \leq (\pi_{\alpha} + \pi_{\beta})|\calX| \cdot d(\bfalpha,\bfbeta)  \; ,
\end{equation}
where
\begin{equation}
\label{eq:dDef}
	d(\bfalpha,\bfzeta) \triangleq \max_{x \in \calX} d(\alpha_x,\zeta_x) \; .
\end{equation}
Before moving on, let us make a few remarks and give some motivation for \eqref{eq:d}, \eqref{eq:DIBound} and \eqref{eq:dDef}. 
First, the usage of ``$\max$" in \eqref{eq:dDef} as opposed to summing over all $x$ is to simplify upcoming derivations. 
Second, recall that we wish to prove the existence of a pair $\alpha,\beta \in \calY$ such that $\DIDeg$ is ``small". 
Then according to \eqref{eq:DIBound}, it suffices to show the existence of a pair that is ``close" in the sense of $d$, assuming that $\pi_{\alpha},\pi_{\beta}$ are also small enough. 
Third, some intuition regarding the need for both $d_1$ and $d_2$ is the following. 
Using $d_2$ alone is not good enough since it diverges in the vicinity of zero. 
Thus, the merger of a pair of close letters with a small entry yields a large value of $d_2$ instead of a small one. 
Using $d_1$ alone, on the other hand, would lead us to a looser bound than desired (see~\cite{Gulcu:DMC}).

Since we are interested in lowering the right hand side of \eqref{eq:DIBound}, we limit our search to a subset of $\calY$, as was done in~\cite{Gulcu:DMC}. Namely,
\[
\calYSmall \triangleq \ppp{y \in \calY: \pi(y)\leq \frac{2}{|\calY|}} \; ,
\]
which implies
\begin{equation}
\label{eq:YSmallSize}
|\calYSmall| \geq \frac{|\calY|}{2} \; . 
\end{equation}
Hence, $\pi_{\alpha}+\pi_{\beta} \leq 4/|\calY|$ and
\begin{equation}
\label{eq:DIBoundYSmall}
\DIDeg \leq \frac{4|\calX|}{|\calY|}\cdot d(\bfalpha,\bfbeta) \; .
\end{equation}

We still need to prove the existence of a pair $\alpha,\beta \in \calYSmall$ that is ``close" in the sense of $d$. 
To that end, as in~\cite{Gulcu:DMC}, we would like to use a sphere-packing approach. 
A typical use of such an argument assumes a proper metric, yet $d$ is not a metric. 
Specifically, the triangle-inequality does not hold:
\begin{multline*}
\underbrace{d\p{\colvec{0.1}{0.9},\colvec{0.2}{0.8}}}_{0.1} + \underbrace{d\p{\colvec{0.2}{0.8},\colvec{0.3}{0.7}}}_{0.05} \\
\ngeq \underbrace{d\p{\colvec{0.1}{0.9},\colvec{0.3}{0.7}}}_{0.2} \; .
\end{multline*}
The absence of a triangle-inequality is a complication that we will overcome, but some care and effort are called for. 
Broadly speaking, as usually done in sphere-packing, we aim to show the existence of a critical ``sphere" radius, $\rcritical=\rcritical(|\calX|,|\calY|) > 0$. 
Such a critical radius will ensure the existence of $\alpha,\beta \in \calYSmall$ for which $d(\bfalpha,\bfbeta) \leq \rcritical$.

\subsection{Non-intersecting ``spheres"}
\begin{figure}[t]	
	\centering
	\begin{tikzpicture}[scale=5]
	\draw[->,line width=1pt] (0,0) -- (0,1cm)  node[left] {$s_1$};
	\draw[->,line width=1pt] (0,0) -- (1cm,0)  node[below] {$s_0$};
	\filldraw[fill=cyan!20,draw=black] (0.28,0.4) rectangle (0.85,0.67) ;
	\fill[black] (0.5cm,0.5cm) circle (0.015cm) ;
	\node at (0.85,0.67) [below left,cyan!20!blue] {\footnotesize $\calB(\bfalpha,r)$} ;
	\draw [line width=1pt] (0.2cm,0.8cm) -- (0.85cm,0.15cm) ;
	\draw[dashed,line width=1pt] (0.1cm,0.9cm) -- (0.2cm,0.8cm) node[pos=0.5,sloped,above] {$s_0+s_1=1$} ;
	\draw[dashed,line width=1pt] (0.85cm,0.15cm) -- (0.95cm,0.05cm)   ;
	\draw[dashed] (0.5cm,0) node[below] {$\alpha_0$}  -- (0.5cm,0.5cm) -- (0,0.5cm) node[left] {$\alpha_1$};
	\draw[dashed] (0.28cm,0) node[below] {}  -- (0.28cm,0.4cm) -- (0,0.4cm) ;
	\draw[dashed] (0.85cm,0) node[below] {}  -- (0.85cm,0.4cm) ;
	\draw[dashed] (0.28cm,0.67cm) -- (0,0.67cm) ;
	\draw[dashed] (0.4cm,0.6cm) -- (0,0.6cm) ;
	\draw [thick, black,decorate,decoration={brace,mirror},yshift=-0.1cm] (0.29,0) -- (0.49,0) node[black,midway,below] {\footnotesize $\omegadown(\alpha_0,r)$};
	\draw [thick, black,decorate,decoration={brace,mirror},yshift=-0.1cm] (0.51,0) -- (0.84,0) node[black,midway,below] {\footnotesize $\omegaup(\alpha_0,r)$};		 
	\draw [thick, black,decorate,decoration={brace},xshift=-0.12cm] (0,0.41) -- (0,0.49) node[black,midway,left] {\footnotesize $\omegadown(\alpha_1,r)$};
	\draw [thick, black,decorate,decoration={brace},xshift=-0.12cm] (0,0.51) -- (0,0.59) node[black,midway,left] {\footnotesize $\omegadown(\alpha_1,r)$};
	\draw [thick, black,decorate,decoration={brace},xshift=-0.4cm] (0,0.51) -- (0,0.66) node[black,midway,left] {\footnotesize $\omegaup(\alpha_1,r)$};
	\draw [thick, red,decorate,decoration={brace},xshift=0.025cm] (0.33,0.67) -- (0.6,0.4) node[red,midway,right] {\footnotesize $\calBK(\bfalpha,r)$};
	\draw [thick, orange!90!black,decorate,decoration={brace,mirror},xshift=-0.025cm] (0.4,0.6) -- (0.6,0.4) node[orange!90!black ] at (0.41cm,0.45cm) {\footnotesize $\calC(\bfalpha,r)$};
	\draw [thick, green!50!black,decorate,decoration={brace,mirror},xshift=0.025cm] (0,0.51) -- (0,0.59) node[green!50!black,midway,right] {\footnotesize $\calQ'(\bfalpha,r)$};	
	\end{tikzpicture} 
	\caption{The sets $\calB(\bfalpha,r)$, $\calBK(\bfalpha,r)$, $\calC(\bfalpha,r)$ and $\calQ'(\bfalpha,r)$ for the binary case ($|\calX|=2$) assuming $\alpha_1 < \alpha_0$ in the $(s_0,s_1)$ plane.}
	\label{fig:spheres}
\end{figure}
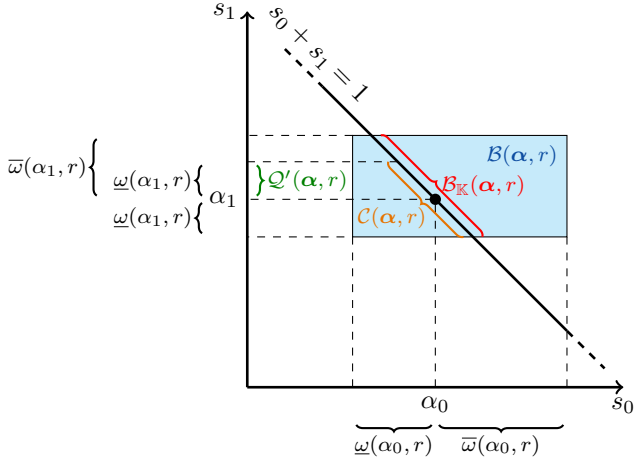

In this section we gradually derive our equivalent of a ``sphere''. 
For convenience, some of the sets defined in the process are illustrated in \Cref{fig:spheres} for the binary case.
We start by giving explicit equations for the ``spheres" corresponding to $d_1$ and $d_2$.

\begin{lemma}
	\label{lm:B1B2}
	For $\nonnegvar \geq 0$ and $r>0$, define the sets $\calB_1(\nonnegvar,r)$ and $\calB_2(\nonnegvar,r)$ as
	\begin{align*}
	\calB_1(\nonnegvar,r) &\triangleq \{\zeta \in \reals : d_1(\nonnegvar,\zeta) \leq r\} \; ,\\
	\calB_2(\nonnegvar,r) &\triangleq \{\zeta \in \reals : d_2(\nonnegvar,\zeta) \leq r\} \; .
	\end{align*}
	Then,
	\[
	\calB_1(\nonnegvar,r) = \{\zeta \in \reals : -r \leq \zeta-\nonnegvar \leq r \} \; ,
	\]
	and
	\[
	\calB_2(\nonnegvar,r) = \{\zeta \in \reals :  - \sqrt{\frac{r^2}{4} + \nonnegvar \cdot r} + \frac{r}{2} \leq \zeta-\nonnegvar  \leq \sqrt{\nonnegvar \cdot r} \} \; .
	\]
\end{lemma}
\begin{IEEEproof}
	Assume $\zeta \in \calB_1(\nonnegvar,r)$. 
	Then $\zeta$ satisfies $|\zeta-\nonnegvar| \leq r$, which is equivalent to $-r \leq \zeta - \nonnegvar \leq r$, and we got the desired result for $\calB_1(\nonnegvar,r)$. 
	Assume now $\zeta \in \calB_2(\nonnegvar,r)$. 
	If $\zeta \geq \nonnegvar$, then $\min(\nonnegvar,\zeta)=\nonnegvar$, and thus
	\[
	\frac{(\zeta-\nonnegvar)^2}{\nonnegvar} \leq r \; ,
	\]
	which implies
	\begin{equation}
	\label{eq:d2Right}
	0 \leq \zeta - \nonnegvar \leq \sqrt{\nonnegvar \cdot r} \; .
	\end{equation}
	If $\zeta \leq \nonnegvar$, then $\min(\nonnegvar,\zeta)=\zeta$, and thus,
	\[
	\frac{(\zeta-\nonnegvar)^2}{\zeta} \leq r \; ,
	\]
	which implies
	\begin{equation}
	\label{eq:d2Left}
	 - \sqrt{\frac{r^2}{4} + \nonnegvar \cdot r} + \frac{r}{2} \leq \zeta-\nonnegvar  \leq 0 \; .
	\end{equation}
	The union of \eqref{eq:d2Right} and \eqref{eq:d2Left} yields the desired result for $\calB_2(\nonnegvar,r)$.
\end{IEEEproof}

Thus, we define
\[
\calB(\nonnegvar,r) \triangleq \{\zeta \in \reals : d(\nonnegvar,\zeta) \leq r\} \; ,
\]
and note that,
\[
\calB(\nonnegvar,r)=\calB_1(\nonnegvar,r) \cup \calB_2(\nonnegvar,r) \; ,
\]
since $d$ takes the $\min$ of the two distances. 
Namely,
\begin{equation}
\label{eq:calBscalar}
\calB(\nonnegvar,r) = \left\{\zeta \in \reals :  -\omegadown(\nonnegvar,r) \leq \zeta-\nonnegvar \leq \omegaup(\nonnegvar,r) \right\} \; ,
\end{equation}
where
\begin{align}
\label{eq:omegaUpDown}
\begin{split}
\omegadown(\nonnegvar,r) &\triangleq \max\left(\sqrt{\frac{r^2}{4} + \nonnegvar \cdot r} - \frac{r}{2},r\right) \\
&= \begin{cases}
	\sqrt{\frac{r^2}{4} + \nonnegvar \cdot r} - \frac{r}{2} & \nonnegvar \geq 2r \; , \\
	r & \nonnegvar \leq 2r \; ,
	\end{cases}  \\
\omegaup(\nonnegvar,r) &\triangleq  \max\left(\sqrt{\nonnegvar \cdot r},r\right)  \\
&= \begin{cases}
	\sqrt{\nonnegvar \cdot r} & \nonnegvar \geq r \; , \\
	r & \nonnegvar \leq r \; .
	\end{cases}	
\end{split}
\end{align}
To extend $\calB$ to vectors we define $\realsX$ as the set of vectors with real entries that are indexed by $\calX$,
\[
\realsX \triangleq \ppp{\bfzeta=(\zeta_x)_{x\in\calX}:\zeta_x \in \reals} \; .
\]
The set $\realsKX$ is defined as the set of vectors from $\realsX$ with entries summing to $1$,
\[
\realsKX \triangleq \ppp{\bfzeta \in \realsX: \sum_{x\in\calX} \zeta_x=1} \; .
\]
The set $\realsKplusX$ is the set of probability vectors. 
Namely, the set of vectors from $\realsKX$ with non-negative entries,
\[
\realsKplusX \triangleq \ppp{\bfzeta \in \realsKX: \zeta_x \geq 0} \; .
\]
We can now define $\calB(\bfalpha,r)$. For $\bfalpha \in \realsKplusX$ let
\begin{equation}
\label{eq:calBVectorDef}
\calB(\bfalpha,r) \triangleq \ppp{\bfzeta \in \realsX:d(\bfalpha,\bfzeta) \leq r } \; .
\end{equation}
Using \eqref{eq:dDef} and \eqref{eq:calBscalar} we have a simple characterization of $\calB(\bfalpha,r)$ as a box: a Cartesian product of segments. 
That is, 
\begin{multline}
\label{eq:calBVector}
\calB(\bfalpha,r) = \Big\{ \bfzeta \in \realsX :  \\
-\omegadown(\alpha_x,r) \leq \zeta_x-\alpha_x \leq \omegaup(\alpha_x,r) \Big\} \; .
\end{multline}
We stress that the box $\calB(\bfalpha,r)$ contains $\bfalpha$, but is not necessarily centered at it.

Recall our aim is finding an $\rcritical$. 
Using our current notation, $\rcritical$ must imply the existence of distinct $\alpha,\beta \in \calYSmall$ such that $\bfbeta \in \calB(\bfalpha,\rcritical)$. 
Naturally, in light of \eqref{eq:DIBound},  we aim to pick $\rcritical$ as small as we can.

Note that $\calB(\bfalpha,r)$ is a set of vectors in $\realsX$. 
However, since the boxes are induced by points $\bfalpha$ in the subspace $\realsKplusX$ of $\realsX$, the sphere-packing would yield a tighter result if performed in $\realsKX$ rather than in $\realsX$. 
Then, for $\bfalpha \in \realsKplusX$ and $r>0$, let us define
\begin{equation}
\label{eq:calBK}
\calBK(\bfalpha,r) = \calB(\bfalpha,r) \cap \realsKX \; .
\end{equation}
When considering $\calBK(\bfalpha,r)$ in place of $\calB(\bfalpha,r)$, we have gained in that the affine dimension (see \cite[Section 2.1.3]{Boyd:Convex})  of $\calBK(\bfalpha,r)$ is $|\calX|-1$ while that of $\calB(\bfalpha,r)$ is $|\calX|$. 
However, we have lost in simplicity: the set $\calBK(\bfalpha,r)$ is not a box. 
Indeed, a moment's thought reveals that any subset of $\realsKX$ with more than one element cannot be a box.

We now show how to overcome the above loss. 
That is, we show a subset of $\calBK(\bfalpha,r)$ which is --- up to a simple transform --- a box. 
Denote the index of the largest entry of a vector $\bfalpha \in \realsKX$ as $\xmax(\bfalpha)$, namely,
\[
\xmax(\bfalpha) \triangleq \argmax_{x \in \calX} \alpha_x \; .
\]
In case of ties, define $\xmax(\bfalpha)$ in an arbitrary yet consistent manner. 
For $\xmax = \xmax(\bfalpha)$ given, or clear from the context, define $\bfzeta'$ as $\bfzeta$, with index $\xmax$ deleted. 
That is, for a given $\bfzeta \in \realsKX$,
\[
\bfzeta' \triangleq (\zeta_x)_{x \in \calX'} \in \realsXminusone \; , 
\]
where $\calX' \triangleq \calX \setminus \{\xmax\}$. 
Note that for $\bfzeta \in \realsKX$, all the entries sum to one. 
Thus, given $\bfzeta'$ and $\xmax$ we know $\bfzeta$. 
Next, for $\bfalpha \in \realsKplusX$ and $r>0$, define the set
\begin{multline}
\label{eq:calC}
\calC(\bfalpha,r) \triangleq \{ \bfzeta \in \realsKX  :\\
 \forall x \in \calX' \; , \;  -\omega'(\alpha_x,r) \leq \zeta_x - \alpha_x \leq \omega'(\alpha_x,r) \} \; , 
\end{multline}
where $\xmax = \xmax(\bfalpha)$ and
\begin{align}
\label{eq:omegaPrime}
\begin{split}
\omega'(\nonnegvar,r) &\triangleq \frac{\omegadown(\nonnegvar,r)}{|\calX|-1} \\
&= \frac{\max\left(\sqrt{r^2/4 + \nonnegvar \cdot r} - r/2,r\right)}{|\calX|-1} \; .
\end{split}
\end{align}

\begin{lemma}
	\label{lemm:CinB}
	Let $\bfalpha \in \realsKplusX$ and $r > 0$ be given. Let $\xmax = \xmax(\bfalpha)$. 
	Then,
	\[
	\calC(\bfalpha,r) \subset \calBK(\bfalpha,r) \; .
	\]
\end{lemma}
\begin{IEEEproof}
	By \eqref{eq:omegaUpDown}, we see that $0 \leq \omegadown(\alpha_x,r) \leq \omegaup(\alpha_x,r)$. 
	Thus, since \eqref{eq:calC} holds, it suffices to show that
	\begin{equation}
	\label{eq:omegaConditionsOnXmax}
	-\omegadown(\alpha_{\xmax},r) \leq \zeta_{\xmax} - \alpha_{\xmax} \leq \omegadown(\alpha_{\xmax},r) \; .
	\end{equation}
	Indeed, summing the condition in \eqref{eq:calC} over all $x \in \calX'$ gives
	\[
	\sum_{x \in \calX'} -\omega'(\alpha_x,r) \leq \sum_{x \in \calX'}\zeta_x -\sum_{x \in \calX'} \alpha_x \leq \sum_{x \in \calX'} \omega'(\alpha_x,r) \; .
	\]
	Since $\omegadown(\alpha_x,r)$ is a monotonically non-decreasing function of $\alpha_x$, we can simplify the above to
	\[
	-\omegadown(\alpha_{\xmax},r) \leq \sum_{x \in \calX'}\zeta_x -\sum_{x \in \calX'} \alpha_x \leq \omegadown(\alpha_{\xmax},r) \; .
	\]
	Since both $\bfzeta$ and $\bfalpha$ are in $\realsKX$, the middle term in the above is $\alpha_{\xmax}- \zeta_{\xmax}$. 
	Thus, \eqref{eq:omegaConditionsOnXmax} follows. 
\end{IEEEproof}

We remind ourselves of our ultimate goal by stating the following corollary, immediate from the previous lemma, together with \eqref{eq:DIBoundYSmall}, \eqref{eq:calBVector} and \eqref{eq:calBK}.
\begin{corollary}
	\label{cor:DIDegBound}
	Let $\alpha,\beta \in \calYSmall$ be such that $\bfbeta \in \calC(\bfalpha,r)$. Then, merging $\alpha$ and $\beta$ induces a penalty $\DIDeg$ of at most
	\[
	\DIDeg \leq \frac{4 |\calX|r}{|\calY|}   \; .
	\]
\end{corollary}

As outlined before, our aim is to find an $\rcritical$ for which the conditions of the above corollary surely hold, for some $\alpha$ and $\beta$. 
A standard sphere-packing approach to finding such an $\rcritical$ is to consider the intersection of spheres of radius $\rcritical/2$. 
Since the triangle inequality does not hold for $d$, we must use a somewhat different approach. 
Towards that end, define the positive quadrant associated with $\bfalpha$ and $r$ as 
\begin{multline*}
\calQ'(\bfalpha,r) \triangleq \{ \bfzeta' \in \realsXminusone  : \\
\forall x \in \calX',\;  0 \leq \zeta_x - \alpha_x \leq \omega'(\alpha_x,r)  \} \; , 
\end{multline*}
where $\xmax = \xmax(\bfalpha)$ and $\omega'(\alpha_x,r)$ is as defined in \eqref{eq:omegaPrime}.

\begin{lemma}
	\label{lm:QSpheresNonintersecting}
	Let $\alpha,\beta \in \calY$ be such that $\xmax(\bfalpha) = \xmax(\bfbeta)$. 
	If $\calQ'(\bfalpha,r)$ and $\calQ'(\bfbeta,r)$ have a non-empty intersection, then $d(\bfalpha,\bfbeta) \leq r$.
\end{lemma}
\begin{IEEEproof}
	By (\ref{eq:calBVectorDef}), (\ref{eq:calBK}) and \Cref{lemm:CinB}, it suffices to prove that $\bfbeta \in \calC(\bfalpha,r)$. 
	Define $\calC'(\bfalpha,r)$ as the result of applying a prime operation on each member of $\calC(\bfalpha,r)$, where $\xmax = \xmax(\bfalpha)$. 
	Hence, we must equivalently prove that $\bfbeta' \in \calC'(\bfalpha,r)$. 
	By \eqref{eq:calC}, we must show that for all $x \in \calX'$,
	\begin{equation}
	\label{eq:alphaxbetaxdistance}
	-\omega'(\alpha_x,r) \leq \beta_x - \alpha_x \leq \omega'(\alpha_x,r) \; .
	\end{equation}
	
	Since we know that the intersection of $\calQ'(\bfalpha,r)$ and $\calQ'(\bfbeta,r)$ is non-empty, let $\bfzeta'$ be a member of both sets. 
	Thus, we know that for $x\in \calX'$,
	\[
	0 \leq \zeta_x - \alpha_x \leq \omega'(\alpha_x,r) \; ,
	\]
	and
	\[
	0 \leq \zeta_x - \beta_x \leq \omega'(\beta_x,r) \; .
	\]
	For each $x\in \calX'$ we must consider two cases: $\alpha_x \leq \beta_x$ and $\alpha_x > \beta_x$.
	
	Consider first the case $\alpha_x \leq \beta_x$. 
	Since $\zeta_x - \alpha_x \leq \omega'(\alpha_x,r)$ and $\beta_x - \zeta_x \leq 0$, we conclude that $\beta_x - \alpha_x \leq \omega'(\alpha_x,r)$. 
	Conversely, since $\beta_x - \alpha_x \geq 0$ and, by \eqref{eq:omegaPrime}, $\omega'(\alpha_x,r) \geq 0$, we have that $\beta_x - \alpha_x \geq - \omega'(\alpha_x,r)$. 
	Thus we have shown that both inequalities in \eqref{eq:alphaxbetaxdistance} hold.
	
	To finish the proof, consider the case $\alpha_x > \beta_x$. 
	We have already established that $\omega'(\alpha_x,r) \geq 0$. 
	Thus, since by assumption $\beta_x - \alpha_x \leq 0$, we have that $\beta_x - \alpha_x \leq \omega'(\alpha_x,r)$. 
	Conversely, since $\zeta_x - \beta_x \leq \omega'(\beta_x,r)$ and $\alpha_x - \zeta_x \leq 0$, we have that $\alpha_x - \beta_x \leq \omega'(\beta_x,r)$. 
	We now recall that by \eqref{eq:omegaPrime}, the fact that $\alpha_x \geq \beta_x$ implies that $\omega'(\beta_x,r) \leq \omega'(\alpha_x,r)$. 
	Thus, $\alpha_x - \beta_x \leq \omega'(\alpha_x,r)$. 
	Negating gives $\beta_x - \alpha_x \geq -\omega'(\alpha_x,r)$, and we have once again proved the two inequalities in \eqref{eq:alphaxbetaxdistance}.
\end{IEEEproof}

\subsection{Weighted ``sphere"-packing}
At first glance, thanks to the above lemma, the quadrant $\calQ'(\bfalpha,r)$ could have been the equivalent of a ``sphere" in the sense of $d$. 
However, recalling \eqref{eq:omegaPrime}, we see that the dimensions of the quadrant $\calQ'(\bfalpha,r)$ are dependent on the point $\bfalpha$. 
Specifically, for $r$ fixed, a larger $\alpha_x$ is preferable. That is, the side length $\omega'(\alpha_x,r)$, corresponding to coordinate $x$, is increasing in $\alpha_x$. 
We now show  how to partially offset this dependence on $\bfalpha$. 
Towards this end,  we define a density over $\realsXminusone$ and derive a lower bound on the weight of a ``sphere" that does not depend on $\bfalpha$.  
Let $\density : \reals \to \reals$ be defined as
\[
\density(\zeta) \triangleq \frac{1}{2\sqrt{\zeta}} \; .
\]
Next, for $\bfzeta' \in \realsXminusone$, abuse notation and define $\density : \realsXminusone \to \reals$ as
\[
\density(\bfzeta') \triangleq \prod_{x \in \calX'} \density(\zeta_x) \; .
\]
The weight of $\calQ'(\bfalpha,r)$ is then defined as
\[
\weightQTag \triangleq \int_{\calQ'(\bfalpha,r)} \density \dd \bfzeta' \; .
\]
The following lemma proposes a lower bound on $\weightQTag$ that does not depend on $\bfalpha$.
\begin{lemma}
	\label{lm:weightQTagLowerBound}
	The weight $\weightQTag$ satisfies
	\begin{equation}
	\label{eq:weightQTagLowerBound}
	\weightQTag \geq r^{\frac{|\calX|-1}{2}}\p{\sqrt{2+\frac{1}{|\calX|-1}}-\sqrt{2}}^{|\calX|-1}  .
	\end{equation}
\end{lemma}
\begin{IEEEproof}
	Since $\density(\bfzeta')$ is a product,
	\begin{align*}
	\weightQTag &= \prod_{x\in\calX'} \int_{\alpha_x}^{\alpha_x+\omega'(\alpha_x,r)} \frac{\mathrm{d}\zeta_x}{2\sqrt{\zeta_x}} \\
	&= \prod_{x\in\calX'} \psi_r(\alpha_x) \; ,
	\end{align*}
	where $\psi_r(\nonnegvar) \triangleq \sqrt{\nonnegvar+\omega'(\nonnegvar,r)}-\sqrt{\nonnegvar}$.
	It suffices to show that $\psi_r(\nonnegvar)$ is decreasing for $\nonnegvar < 2r$, and increasing for $\nonnegvar >2r$. 
	Assume first that $\nonnegvar <2r$. Then,
	\[
	\frac{\dd \psi_r}{\dd \nonnegvar} =\frac{1}{2\sqrt{\nonnegvar+r/(|\calX|-1)}} - \frac{1}{2\sqrt{\nonnegvar}} <0 \; ,	
	\]
	for all $\nonnegvar \geq 0$. 
	Assume now that $\nonnegvar > 2r$. Then,
		\begin{align*}
		\frac{\dd \psi_r}{\dd \nonnegvar}&= \frac{1}{2} \p{ \sqrt{\nonnegvar+\frac{\sqrt{r^2/4+\nonnegvar \cdot r}-r/2}{|\calX|-1}} }^{-1} \\
		&\quad\cdot \p{ 1+\frac{r}{2(|\calX|-1)\sqrt{r^2/4+\nonnegvar \cdot r} }  } - \frac{1}{2\sqrt{\nonnegvar}} \; .
		\end{align*}
	Comparing the derivative to zero and solving for $\nonnegvar$ yields
	\begin{multline*}
		 \nonnegvar+\frac{\sqrt{r^2/4+\nonnegvar \cdot r}-r/2}{|\calX|-1}\\
		 =\nonnegvar \p{ 1+\frac{r}{2(|\calX|-1)\sqrt{r^2/4+\nonnegvar \cdot r} }  }^2 \; ,
	\end{multline*}
	which is equivalent to
		\begin{multline*}
		\frac{\sqrt{r^2/4+\nonnegvar \cdot r}-r/2}{|\calX|-1} = \\
		\frac{\nonnegvar \cdot r}{(|\calX|-1)\sqrt{r^2/4+\nonnegvar \cdot r} } + \frac{\nonnegvar \cdot r^2}{4(|\calX|-1)^2\left( r^2/4+\nonnegvar \cdot r\right)} \; .	
		\end{multline*}
	We define $\xi \triangleq \sqrt{\frac{r^2}{4}+\nonnegvar \cdot r}$ and get
	\[
	\xi - \frac{r}{2} = \frac{\xi^2-\frac{r^2}{4}}{\xi}+\frac{r(\xi^2-\frac{r^2}{4})}{4(|\calX|-1)\xi^2}		\; .
	\]
	Now solving for $\xi$ we have
	\[
	(2|\calX|-1)\xi^2-(|\calX|-1)r\xi-\frac{r^2}{4}=0 \; ,
	\]
	and since $\xi > 0$ the only solution is $\xi=r/2$ which implies $\nonnegvar = 0$. 
	We note that this is not a real solution since the derivative is not defined for $\nonnegvar=0$. 
	Hence, the derivative is either non-negative or non-positive. 
	By plugging $\nonnegvar=2r$ in the derivative we get
	\begin{align*}
	\frac{\dd \psi_r}{\dd \nonnegvar}\bigg|_{\nonnegvar=2r} &= \frac{1}{2} \p{ \sqrt{2r+\frac{r}{|\calX|-1}} }^{-1} \cdot \p{ 1+\frac{r}{(|\calX|-1)3r  }}\\
	&\quad - \frac{1}{2\sqrt{2r}} \\
	&=\frac{1}{2r\sqrt{2}}\p{ \frac{1+\frac{1}{3(|\calX|-1)}}{\sqrt{1+\frac{1}{2(|\calX|-1)}}}     -1} \\
	&\geq 0\; ,
	\end{align*}
	for $|\calX| \geq 2$ and $r>0$. 
	Thus, by continuity of $\psi_r$,
	\[
	\psi_r(\alpha_x) \geq \psi_r(2r) \; , 
	\]
	and since this lower bound does not depend on $\alpha_x$ we get \eqref{eq:weightQTagLowerBound}.
\end{IEEEproof}

We divide the letters in $\calYSmall$ according to their $\xmax$ value to $|\calX|$ subsets (at most). 
The largest subset is denoted by $\calY'$, and we henceforth fix $\xmax$ accordingly. 
We limit our search to $\calY'$. 

Let $\calV'$ be the union of all the quadrants corresponding to possible choices of $\bfalpha$. 
Namely,
\[
\calV' \triangleq \bigcup_{\stackrel{\bfalpha \in \realsKplusX}{x_{\max}(\bfalpha)=x_{\max} } } \calQ'(\bfalpha,r)  \; .
\]
In order to bound the weight of $\calV'$, we introduce the simpler set 
\[
\calU' \triangleq \ppp{ \bfzeta' \in \realsXminusone: \sum_{x\in\calX'} \zeta_x \leq 2, \; \zeta_x \geq 0 \; \forall x\in\calX' } \;.
\]
The constraint $r \leq 1$ in the following lemma will be motivated shortly.
\begin{lemma}
	\label{lm:VSubsetU}
	Let $r \leq 1 $. Then, $\calV' \subseteq \calU'$.
\end{lemma} 
\begin{IEEEproof}
	Assume $\bfzeta'\in\calV'$. 
	Then, there exists $\bfalpha \in \realsKplusX$ such that $0 \leq \zeta_x -\alpha_x \leq \omega'(\alpha_x,r)$ for all $x \in \calX'$. 
	Hence, $\zeta_x \geq 0$  for all $x \in \calX'$. 
	Moreover,
	\begin{align}
	\label{eq:sumOfZetaTag}
	\sum_{x \in \calX'} \zeta_x &\leq \sum_{x \in \calX'} \alpha_x + \sum_{x \in \calX'} \omega'(\alpha_x,r) \nonumber \\
	&\leq 1-\alpha_{x_{\max}} + \omegadown(\alpha_{\xmax},r) \;.
	\end{align}
	There are two cases to consider. 
	In the case where $\alpha_{x_{\max}} \geq 2r$ we have
	\begin{align*}
	\sum_{x \in \calX'} \zeta_x &\leq 1-\alpha_{x_{\max}} + \sqrt{\frac{r^2}{4} + \alpha_{x_{\max}}r} - \frac{r}{2} \\
	&\leq 1-\alpha_{x_{\max}} + \sqrt{\frac{\alpha_{x_{\max}}^2}{16} + \frac{\alpha_{x_{\max}}^2}{2}} - \frac{r}{2} \\
	&=1-\frac{\alpha_{x_{\max}}}{4}-\frac{r}{2}\\
	&\leq 2	\;,
	\end{align*}
	where the second inequality is due to the assumption $\alpha_{x_{\max}} \geq 2r$. 
	In the case where $\alpha_{x_{\max}} \leq 2r$, \eqref{eq:sumOfZetaTag} becomes
	\begin{align*}
	\sum_{x \in \calX'} \zeta_x &\leq 1-\alpha_{x_{\max}} + r \\
	&\leq 2-\alpha_{x_{\max}} \\
	&\leq 2 \; ,
	\end{align*}
	where we assumed $r\leq 1$. 
	Therefore, $\bfzeta' \in \calU'$.
\end{IEEEproof}

The lemma above and the non-negativity of $\varphi$, enable us to upper bound the weight of $\calV'$, denoted by $\weightVTag $, using
\[
\weightVTag \triangleq \int_{\calV'} \density \dd \bfzeta' \leq \int_{\calU'} \density \dd \bfzeta' \; .
\]
We define the mapping $\rho_x = \sqrt{\zeta_x}$ for all $x \in \calX'$ and perform a change of variables. 
As a result, $\calU'$ is mapped to
\[
\calS' \triangleq \ppp{ \bfrho'  \in \realsXminusone: \sum_{x \in \calX'} \rho_x^2 \leq 2, \; \rho_x \geq 0 } \; ,
\]
which is a quadrant of a $|\calX|-1$ dimensional ball of a $\sqrt{2}$ radius. 
The density function $\varphi$ transforms into the unit uniform density function since
\[
\frac{\dd \zeta_x}{2\sqrt{\zeta_x}} = \dd \rho_x \; .
\]
Hence,
\begin{align}
\label{eq:VWeight}
\weightVTag &\leq \int_{\calS'} \dd V \nonumber \\
&= \frac{1}{2^{|\calX|-1}} \frac{\pi^{\frac{|\calX|-1}{2}}}{\Gamma\p{1+\frac{|\calX|-1}{2}}} 2^{\frac{|\calX|-1}{2}} \nonumber \\
&=\left(\frac{\pi}{2}\right)^{\frac{|\calX|-1}{2}} \frac{1}{\Gamma\p{1+\frac{|\calX|-1}{2}}} \; ,
\end{align}
where we have used the well known expression for the volume of a multidimensional ball. 
Thus, we are ready to prove \Cref{thm:DIDeg}.
\begin{IEEEproof}[Proof of \Cref{thm:DIDeg}]
	Recall that we are assuming $|\calY| >2|\calX|$. 
	According to the definition of $\calY'$, we get
	\begin{equation}
	\label{eq:calYTagSize}
	|\calY'| \geq \frac{|\calYSmall|}{|\calX|} \geq \frac{|\calY|}{2|\calX|} > 1 \; ,
	\end{equation}
	where we used \eqref{eq:YSmallSize}. 
As a result, we have at least two points in $\calY'$, and are therefore in a position to apply a sphere-packing argument. 
Towards this end, let $r$ be such that the starred equality in the following derivation holds:
\begin{align}
	\begin{split}
		\label{eq:spherePackingDerivation}
		\sum_{\bfalpha \in \calY'} &\weightQTag  \\
		&\geq \frac{|\calY|}{2|\calX|}\cdot r^{\frac{|\calX|-1}{2}}\p{\sqrt{2+\frac{1}{|\calX|-1}}-\sqrt{2}}^{|\calX|-1} \\
		&\stackrel{(\ast)}{=} \p{\frac{\pi}{2}} ^{\frac{|\calX|-1}{2}} \frac{1}{\Gamma\p{1+\frac{|\calX|-1}{2}}} \\
		&\geq \weightVTag \; .
	\end{split}
\end{align}
Namely,
\begin{multline}
	\label{eq:rstarDef}
	r \triangleq \frac{\pi}{4}\p{\sqrt{1+\frac{1}{2(|\calX|-1)}}-1}^{-2} \\
	\cdot \p{\frac{2|\calX|}{\Gamma\p{1+\frac{|\calX|-1}{2}}}}^{\frac{2}{|\calX|-1}}  \cdot  |\calY|^{-\frac{2}{|\calX|-1}} \; .
\end{multline}
There are two cases to consider. 
If $r \leq 1$, then all of (\ref{eq:spherePackingDerivation}) holds, by \eqref{eq:weightQTagLowerBound}, \eqref{eq:VWeight} and \eqref{eq:calYTagSize}. 
We take $\rcritical = r$, and deduce the existence of a pair $\alpha,\beta \in \calY'$ for which $d(\bfalpha,\bfbeta) \leq r$. 
Indeed, assuming otherwise would contradict (\ref{eq:spherePackingDerivation}), since each $\calQ'$ in the sum is contained in $\calV'$, and, by \Cref{lm:QSpheresNonintersecting} and our assumption, all summed $\calQ'$ are disjoint.

We next consider the case $r > 1$. 
Now, any pair of letters $\alpha,\beta \in \calY'$ satisfies $d(\bfalpha,\bfbeta) \leq r$. 
Indeed, by \eqref{eq:dScalarDef} and \eqref{eq:dDef},
\[
d(\bfalpha,\bfbeta) \leq \|\bfalpha-\bfbeta\|_{\infty} \leq 1 < r \; ,
\]
where $\|\cdot\|_{\infty}$ is the maximum norm.

We have proved the existence of $\alpha,\beta \in \calY' \subset \calYSmall$ for which $d(\bfalpha,\bfbeta) \leq r$. 
By  (\ref{eq:DIBoundYSmall}) and (\ref{eq:rstarDef}), the proof is finished.
\end{IEEEproof}
Finally, we prove \Cref{thm:DIDegStar}.
\begin{IEEEproof}[Proof of \Cref{thm:DIDegStar}]
	If $L \geq |\calY|$, then obviously $\DIDegStar=0$ which is not the interesting case. 
	If $2|\calX| \leq L <|\calY|$, then applying \Cref{thm:DIDeg} repeatedly $|\calY|-L$ times yields
	\begin{align*}
		\DIDegStar &\leq \sum_{\ell=L+1}^{|\calY|} \mu(|\calX|) \cdot \ell^{-\frac{|\calX|+1}{|\calX|-1}} \\
		&\leq \mu(|\calX|)\int_L^{|\calY|} \ell^{-\frac{|\calX|+1}{|\calX|-1}} \dd \ell \\
		&= \nu(|\calX|) \p{ L^{-\frac{2}{|\calX|-1}}- |\calY|^{-\frac{2}{|\calX|-1}}} \\
		&\leq \nu(|\calX|) \cdot  L^{-\frac{2}{|\calX|-1}} \; ,
	\end{align*}
	by the monotonicity of $\ell^{-(|\calX|+1)/(|\calX|-1)}$. 
	The bound is tight in the power-law sense by \cite[Theorem 2]{Tal:ModerateSizes}, in which a sequence of channels is proved to obtain 
	\[
	\DIDegStar \geq \delta(|\calX|) \cdot L^{-\frac{2}{|\calX|-1}} \; ,
	\]
	for a specified $\delta(\cdot)$.
\end{IEEEproof}

We note that \Cref{thm:DIDegStar} can be generalized to channels with a continuous output alphabet. 
This is done using arbitrarily close approximating degraded channels~\cite{Tal:ConstructingPolarCodes}\cite{PeregTal:Upgrading}, with corresponding large finite output alphabets.

\subsection{Symmetric channels}
\label{subsec:symmetricChannels}
We note that degrading a symmetric channel optimally does not necessarily yield a symmetric channel~\cite[Theorem 2]{Koch:AsymmetricQuantizers}\cite{Alirezaei:AsymmetricQuantizers}.
However, often, the fact that the resultant channel is symmetric is important. 
For example, \cite[Theorem 4]{Arikan:PolarCodes} proves that if the underlying binary-input channel is symmetric, then in many respects, any choice of frozen bits is as good as any other. 
Thus, we would like an analog of Theorem~\ref{thm:DIDegStar}, for the case in which all channels involved are symmetric. 
We have indeed found such an analog, for a restricted set of symmetric channels, defined as \emph{cyclo-symmetric} channels.

A channel $W:\calX \rightarrow \calY$ is cyclo-symmetric if the following holds.
\begin{enumerate}
\item The input alphabet is labeled $\calX = \{0,1,\ldots,|\calX|-1\}$.
\item The output alphabet size is a multiple of $|\calX|$, and partitioned into $|\calY|/|\calX|$ disjoint sets $\{\calY_i\}_{i=1}^{|\calY|/|\calX|}$. Each such $\calY_i$ contains $|\calX|$ members,
\[
\calY_i = \left\{ y_i^{(0)}, y_i^{(1)}, \ldots, y_i^{(|\calX|-1)} \right\} \; .
\]
\item For $0 \leq \theta \leq |\calX|-1$, 
\begin{equation}
\label{eq:cycloSymmetry}
W(y_i^{(0)}|x) = W(y_i^{(\theta)}|x+\theta) \; ,
\end{equation}
where $x+\theta$ is short for $x+\theta \mod |\calX|$.
\end{enumerate}
Note that a cyclo-symmetric channel is symmetric, according to the definition in \cite[page 94]{Gallager:IT}. 
Hence, the capacity-achieving input distribution is the uniform input distribution $\pi(x)=1/|\calX|$ \cite[Theorem 4.5.2]{Gallager:IT}. 
We remark in passing that in the binary-input case, $|\calX|=2$, a symmetric channel is essentially cyclo-symmetric as well: the only problematic symbol is the erasure symbol, which can be split, as discussed in \cite[Lemma 4]{Tal:Construct}.

\begin{theorem}
	\label{theo:symmetric}
	Let a DMC $W:\calX \rightarrow \calY$ be cyclo-symmetric and satisfy $|\calY|>2|\calX|$ and let $L \geq 2|\calX|$ be a multiple of $|\calX|$. 
	Fix the input distribution to be uniform, $\pi(x) = 1/|\calX|$ for all $x \in \calX$. 
	Then,
	\[
	\DIDegStar =\min_{\substack{{Q,\Phi:Q \preccurlyeq W,} \\ {|Q| \leq L}} } I(W)-I(Q)= O\p{L^{-\frac{2}{|\calX|-1}}} \; ,
	\] 
        where the optimization is over $Q$ that are cyclo-symmetric.
	In particular, 
	\[
	\DIDegStar \leq \nu(|\calX|)\cdot L^{-\frac{2}{|\calX|-1}} \; , 
	\]
	where $\nu(|\calX|)$ is as defined in Theorem~\ref{thm:DIDegStar}. This bound is attained by a simple modification of greedy-merge, and is tight in the power-law sense.
\end{theorem}

Before getting into the proof, let us explain the modification of greedy-merge mentioned in the theorem. 
Using the above notation, in greedy-merge we are to choose the $y_i^{(t)}$ and $y_j^{(t')}$ whose merger results in the smallest drop in mutual information between input and output. 
In our modified algorithm, we limit our search to the case in which $i \neq j$. 
Namely, the symbols are taken from $\calY_i$ and $\calY_j$, and these sets are distinct. 
After we have completed our search and found the above $y_i^{(t)}$ and $y_j^{(t')}$, we merge $\calY_i$ and $\calY_j$ into a new set $\calY_{ij}$, in the following sense: for $0 \leq \theta \leq |\calX|-1$, we merge $y_i^{(t+\theta)}$ and $y_j^{(t'+\theta)}$ into $y_{ij}^{(\theta)}$, where $t+\theta$ and $t'+\theta$ are calculated modulo $|\calX|$. 
As in greedy-merge, the operation of merging $\calY_i$ and $\calY_j$ is repeated until the output alphabet size is small enough.
\begin{IEEEproof}[Proof of Theorem~\ref{theo:symmetric}]
We start by proving that the resulting channel $Q$ is cyclo-symmetric. 
To do so, we prove that each merging iteration  --- merging of sets $\calY_i$ and $\calY_j$ --- preserves cyclo-symmetry. 
Suppose for notational convenience that only one such merge iteration is needed, taking us from channel $W$ to channel $Q$. 
Let the merging be carried out using the notation above: $\calY_i$ and $\calY_j$ are merged to form $\calY_{ij}$, with $y_i^{(t)}$ and $y_j^{(t')}$ as the pair initiating the merger. 
To prove that cyclo-symmetry is preserved, we must show that (\ref{eq:cycloSymmetry}) holds. 
Namely, for all $x \in \calX$,
\[
Q(y_{ij}^{(0)}|x) = Q(y_{ij}^{(\theta)}|x+\theta) \; .
\]
The above is equivalent to showing that
\[
W(y_i^{(t)}|x) + W(y_j^{(t')}|x) = W(y_i^{(t+\theta)}|x+\theta) + W(y_j^{(t'+\theta)}|x + \theta) \; ,
\]
which follows immediately from (\ref{eq:cycloSymmetry}). 
Obviously, if the output alphabet of $W$ is a multiple of $|\calX|$, then the output alphabet of $Q$ is smaller by $|\calX|$, and is thus still a multiple of $|\calX|$.

We now move to proving the upper-bound on the difference in mutual information. 
Since Theorem~\ref{thm:DIDegStar} is a direct consequence of Theorem~\ref{thm:DIDeg}, it suffices to prove that each sub-merger of $y_i^{(t+\theta)}$ and $y_j^{(t'+\theta)}$ attains the bound in Theorem~\ref{thm:DIDeg}. 
Namely, the bound corresponding to $\theta = 0,1,\ldots,|\calX|-1$ must hold with respect to $|\calY| - \theta$ output letters.

Let us first consider the case $\theta=0$. 
Recall that for $\theta=0$, the only constraint imposed by our version of greedy-merge is that the two symbols merged, $y_i^{(t)}$ and $y_j^{(t')}$, must have $i \neq j$. 
Apart from this, as in regular greedy-merge, we pick the pair $y_i^{(t)}$ and $y_j^{(t')}$ for which the reduction in mutual information is minimal. 
Thus, we must show that this added constraint is still compatible with the proof of Theorem~\ref{thm:DIDeg}. 
Recall that in the proof of Theorem~\ref{thm:DIDeg}, only symbols with the same $\xmax$ are considered. 
Thus, the proof will indeed be compatible with our version of greedy-merge if we manage to show that all the symbols in a generic subset $\calY_i$ have distinct $\xmax$. 
Indeed, by (\ref{eq:cycloSymmetry}), the $\xmax$ corresponding to $y_i^{(\theta)}$ is simply the $\xmax$ corresponding to $y_i^{(0)}$, shifted by $\theta$ places (where, if needed, ties are broken accordingly).

Recall that we are considering (\ref{eq:DIDegOrder}), with $|\calY|$ replaced by $|\calY| - \theta$. 
Let us term this (\ref{eq:DIDegOrder})'. 
We have just proved that (\ref{eq:DIDegOrder})' holds for $\theta = 0$, and our aim now is to prove it for $1 \leq \theta < |\calX|$. 
Since the input distribution is uniform, we have by (\ref{eq:cycloSymmetry}) that the difference in mutual information between input and output resulting from merging $y_i^{(t+\theta)}$ and $y_j^{(t'+\theta)}$ equals that from merging $y_i^{(t)}$ and $y_j^{(t')}$. 
That is, the LHS of (\ref{eq:DIDegOrder})' is independent of $\theta$. 
Since the RHS of (\ref{eq:DIDegOrder})' is increasing in $\theta$, and the claim holds for $\theta = 0$, we are done.

Lastly, we must prove the claim of tightness, in the power-law sense. 
This is so, since the channels in \cite[Theorem 2]{Tal:ModerateSizes} are essentially cyclo-symmetric. 
That is, consider the output symbols in such a channel. 
All symbols having a corresponding posterior probability vector with period $|\calX|$ can be grouped into subsets satisfying (\ref{eq:cycloSymmetry}). 
The remaining symbols (a vanishing fraction) will have posterior probability vectors with period dividing $|\calX|$. 
These can be ``split'', similarly to how erasure symbols are split in \cite[Lemma 4]{Tal:Construct}, to yield an equivalent channel which is cyclo-symmetric.
 \end{IEEEproof}

\section{Lower bound on upgrading cost}
\label{sec:LBUC}
Recall the definition of upgrading-cost given in (\ref{eq:DIUpgStar}) and (\ref{eq:UCDef}). 
In this section, we derive a lower bound on the upgrading cost.
\begin{theorem}
	\label{thm:DIUpg}
	Given $|\calX|$ and $L$, the upgrading cost defined in \eqref{eq:UCDef} satisfies
	\begin{equation}
	\label{eq:UCLB}
	\UC \geq \kappa(|\calX|) \cdot L^{-\frac{2}{|\calX|-1}} \; ,
	\end{equation}
	where
	\[
	\kappa(|\calX|) \triangleq \frac{|\calX|-1}{2\pi(|\calX|+1)} \cdot \p{\frac{\Gamma\p{1+\frac{|\calX|-1}{2}}}{(|\calX|-1)!}}^{\frac{2}{|\calX|-1}} \; ,
	\]
	and $\Gamma(\cdot )$ is the Gamma function, defined in (\ref{eq:Gamma}). 
\end{theorem}
Note that for large values of $|\calX|$ the Stirling approximation can be applied to simplify $\kappa(|\calX|)$ to
\[
\kappa(|\calX|) \approx \frac{e}{4\pi (|\calX|-1)} \; .
\]

The proof of the above theorem will rely on a sequence of channels that are ``hard'' to upgrade. 
It turns out that these channels are exactly the channels that \cite{Tal:ModerateSizes} proved were hard to \emph{degrade}. 
In fact, more is true: the lower bound of Theorem~\ref{thm:DIUpg} is exactly equal to the lower bound proved in \cite[Theorem~2]{Tal:ModerateSizes}. 
As a result, this section will be rather short: we will first prove two lemmas which are specific to the upgrading case, and then use them to show that a key part of the proof of \cite[Theorem~2]{Tal:ModerateSizes} is applicable to our setting.

We now fix some notation. 
Let $W:\calX \rightarrow \calY$ and $Q:\calX \rightarrow \calZ$ be two DMCs such that $Q$ is upgraded with respect to $W$, that is $W \preccurlyeq Q$. 
We assume, again, without loss of generality, that $\calX,\calY$ and $ \calZ$ are disjoint and that an input distribution is fixed. 
We can thus abuse notation and define
\begin{align}
	\label{eq:yxzx}
	\begin{split}
	y_x &\triangleq W(x|y) \; ,\\
	z_x &\triangleq Q(x|z) \; ,
	\end{split}
\end{align}
and the corresponding vectors $\bfy \triangleq \p{y_x}_{x \in \calX}$, $\bfz \triangleq \p{z_x}_{x \in \calX}$. 

Let us think of $\calY$ as a ``large'' alphabet, that is reduced to a ``small'' alphabet $\calZ$. 
For each $z \in \calZ$, we define $\calA_z$ as the set of letters in $\calY$ that are closest to $z$, with respect to Euclidean distance between posterior vectors. 
That is, for $z \in \calZ$
\begin{equation}
\label{eq:Az}
\calA_z \triangleq \ppp{y \in \calY: z=\argmin_{z' \in \calZ} \norm{\bfz'-\bfy}_2^2 } \; ,
\end{equation}
where $\norm{\cdot}_2$ is the Euclidean norm. 
We stress that the sets $\{\calA_z\}_{z \in \calZ}$ are disjoint. 
That is, ``$\argmin$'' ties are broken in a consistent yet arbitrary manner. 

We now show how the sets $\calA_z$ can be used to derive a lower bound on the cost of upgrading $W$ to $Q$. 
As before, we use the shorthand $\pi_y$ to denote $\pi(y)$.
\begin{lemma}
	\label{lm:DIUpgLB1}
	Let the DMCs $W:\calX \rightarrow \calY$ and $Q:\calX \rightarrow \calZ$ satisfy $W \preccurlyeq Q$. 
	Assume a fixed input distribution. 
	Then,
	\[
	\DIUpg \triangleq I(Q)-I(W) \geq \sum_{z \in \calZ} \Delta(\calA_z) \; ,
	\]
	where
	\begin{equation}
	\label{eq:DeltaAz}
	\Delta(\calA_z) \triangleq \frac{1}{2}\sum_{y \in \calA_z} \pi_y \norm{\bfz-\bfy}_2^2 \; .
	\end{equation}
\end{lemma}
\begin{IEEEproof}
	Using our notation,
	\begin{equation}
	\label{eq:DIUpgH}
	\DIUpg =  \sum_{y \in \calY} \pi_y h(\bfy) - \sum_{z \in \calZ} \pi_z h(\bfz) \; ,
	\end{equation}
	where
	\[
	h(\bfy) \triangleq \sum_{x \in \calX} \eta(y_x) \; ,
	\]
	and $\eta(\cdot)$ was defined in \eqref{eq:etaDef}. 
	Since $W \preccurlyeq Q$, there exists an intermediate channel $\Phi: \calZ \rightarrow \calY$ such that concatenating $\Phi$ to $Q$ results in $W$. 
	We now claim that this concatenation applies also to posterior probabilities,
	\begin{equation}
	\label{eq:bfy}
	\bfy = \sum_{z \in \calZ} \Phi_{z|y} \bfz \; ,
	\end{equation}
	where for $y \in \calY$ and $z \in \calZ$
\begin{equation}
\label{eq:PhiReverse}
\Phi_{z|y} \triangleq \Phi(z|y) \triangleq \frac{\Phi(y|z) \cdot \pi_z}{\pi_y}
\end{equation}
 is the ``reverse'' or ``posterior'' channel, often also called the ``test'' channel. Note that \eqref{eq:bfy} follows from \eqref{eq:Degrading} and an application of Bayes' rule.
	Moreover, by (\ref{eq:PhiReverse}),
	\begin{equation}
	\label{eq:piz}
	\pi_z = \sum_{y \in \calY} \Phi_{z|y} \pi_y \; .
	\end{equation}
	Plugging \eqref{eq:bfy} and \eqref{eq:piz} in \eqref{eq:DIUpgH} yields
	\begin{equation}
	\label{eq:DIUpgPhi}
	\DIUpg = \sum_{y \in \calY} \pi_y \p{h\p{\sum_{z \in \calZ} \Phi_{z|y} \bfz}   - \sum_{z \in \calZ} \Phi_{z|y} h\p{\bfz}  } \; .
	\end{equation}

	 It easily follows from (\ref{eq:PhiReverse}) that $\sum_{z \in \calZ} \Phi_{z|y}=1$. 
	 Hence, since $h(\cdot)$ is concave, we can apply Jensen's inequality to the expression contained by the outer parenthesis of (\ref{eq:DIUpgPhi}) and conclude that it is non-negative. 
	 However, as in  \cite[Corollary~5]{Tal:ModerateSizes}, we invoke a stronger inequality, known as H\"{o}lder's defect formula~\cite[Page~94]{Steele:CS}. 
	 This yields
	\begin{multline*}
	h\p{\sum_{z \in \calZ} \Phi_{z|y} \bfz}   - \sum_{z \in \calZ} \Phi_{z|y} h\p{\bfz}  \geq \\
	 \frac{1}{2} \lambda_{\min} (-\nabla^2 h) \sum_{z \in \calZ} \Phi_{z|y} \Big\|\bfz-\sum_{z' \in \calZ} \Phi_{z'|y} \bfz'\Big\|_2^2 \; ,
	\end{multline*} 
	where $-\nabla^2 h$ is the negated Hessian matrix of $h$, and $\lambda_{\min} (-\nabla^2 h)$ is its smallest eigenvalue. 
	Using \eqref{eq:bfy} for the term inside the norm and $\lambda_{\min} (-\nabla^2 h) \geq 1$ (proved in~\cite[Corollary~5]{Tal:ModerateSizes}), we get
	\[
	h\p{\sum_{z \in \calZ} \Phi_{z|y} \bfz}   - \sum_{z \in \calZ} \Phi_{z|y} h\p{\bfz}  \geq \frac{1}{2} \sum_{z \in \calZ} \Phi_{z|y} \norm{\bfz- \bfy}_2^2 \; .
	\]
	Thus, by (\ref{eq:DIUpgPhi}) and the above,
	\begin{align*}
	\DIUpg &\geq \frac{1}{2} \sum_{y \in \calY} \pi_y \sum_{z \in \calZ} \Phi_{z|y} \norm{\bfz- \bfy}_2^2 \\
	&\geq \frac{1}{2} \sum_{y \in \calY} \pi_y \sum_{z \in \calZ} \Phi_{z|y} \p{\min_{z' \in \calZ}\norm{\bfz'- \bfy}_2^2} \\
	&= \frac{1}{2} \sum_{y \in \calY} \pi_y \min_{z' \in \calZ}\norm{\bfz'- \bfy}_2^2 \; .
	\end{align*}
        Recall that the sets $\ppp{\calA_z}_{z \in \calZ}$ partition $\calY$. 
        Thus, continuing the above,

	\begin{align*}
	\DIUpg &\geq \frac{1}{2} \sum_{z \in \calZ} \sum_{y \in \calA_z} \pi_y \min_{z' \in \calZ}\norm{\bfz'- \bfy}_2^2 \\
	& =  \frac{1}{2} \sum_{z \in \calZ} \sum_{y \in \calA_z} \pi_y \norm{\bfz- \bfy}_2^2 \\
	& =  \sum_{z \in \calZ} \Delta(\calA_z) \; ,
	\end{align*}
where the first and second equalities follow from (\ref{eq:Az}) and (\ref{eq:DeltaAz}), respectively.
\end{IEEEproof}

To coincide with the proof in~\cite[Theorem 2]{Tal:ModerateSizes} we will further lower bound $\DIUpg$ by making use of the following lemma.
\begin{lemma}
	\label{lm:DIUpgLB2}
	Let $\Delta(\calA_z)$ be as defined in \Cref{lm:DIUpgLB1}. 
	Then,
	\[
	\Delta(\calA_z) \geq \tilde{\Delta}(\calA_z) \; ,
	\]
	where
	\begin{equation}
	\label{eq:DeltaTildeDef}
	\tilde{\Delta}(\calA_z) \triangleq \frac{1}{2}\sum_{y \in \calA_z} \pi_y \norm{\bfy-\bfybar_z}_2^2 \; ,
	\end{equation}
	and $\bfybar_z$ is the weighted center of $\calA_z$,
	\[
	\bfybar_z \triangleq \frac{\sum_{y \in \calA_z} \pi_y \bfy}{\sum_{y \in \calA_z} \pi_y} \; .
	\]
\end{lemma}
\begin{IEEEproof}
	Define the function 
	\[
	V(\bfu) \triangleq \frac{1}{2}\sum_{y \in \calA_z} \pi_y \norm{\bfy-\bfu}_2^2 \; .
	\]
	Since the $\pi_y$ are non-negative, $V(\bfu)$ is easily seen to be convex in $\bfu$. 
	Thus, the minimum is calculated by differentiating with respect to $\bfu$ and equating to $0$. 
	Since $V(\bfu)$ is quadratic in $\bfu$, we have a simple closed-form solution,
	\[
	\bfu = \frac{\sum_{y \in \calA_z} \pi_y \bfy}{\sum_{y \in \calA_z} \pi_y} = \bfybar_z \; ,
	\]
	and the proof is finished.
\end{IEEEproof}

We return to proving the main theorem of this section.
\begin{IEEEproof}[Proof of \Cref{thm:DIUpg}]
	According to~\cite[Claim 1]{Tal:ModerateSizes}, a DMC $W:\calX \to \calY$ is optimally degraded to a channel $Q:\calX \to \calZ$ by partitioning $\calY$ to $|\calZ|$ disjoint subsets, denoted by $\{A_z\}_{z \in \calZ}$, and merging all the letters in each subset. 
	It is then shown in~\cite[Corollary~5]{Tal:ModerateSizes} that the loss in mutual information, as a result of this operation, can be lower bounded by $\sum_{z \in \calZ} \tilde{\Delta}(A_z)$, where $\tilde{\Delta}(A_z)$ is defined as in~(\ref{eq:DeltaTildeDef}). 
	As a final step, in~\cite[Section~V]{Tal:ModerateSizes} a specific sequence of channels and input distributions is introduced and analyzed.
	For this sequence, $\sum_{z \in \calZ} \tilde{\Delta}(A_z)$ is lower bounded by the same bound as in~(\ref{eq:UCLB}).

	In our case, as a result of \Cref{lm:DIUpgLB1} and \Cref{lm:DIUpgLB2}, 
	\begin{equation}
	\label{eq:DIUpgDeltaTilde}
	\DIUpg \geq \sum_{z \in \calZ} \tilde{\Delta}(\calA_z) \; .
	\end{equation}
	and  we got the same expression as in~\cite[Corollary 5]{Tal:ModerateSizes}.
	From this point on, the proof is identical to~\cite[Section~V]{Tal:ModerateSizes}.	
\end{IEEEproof}

\section{Optimal binary upgrading}
\label{sec:OBU}
\subsection{Main result}
In this section we show an efficient algorithmic implementation of optimal upgrading for the BDMC case, $|\calX|=2$. 
As in the degrading case \cite{Kurkoski:Quantization}, the algorithm will be an instance of dynamic programming. 
The following theorem is cardinal, and is the main result of the section.
\begin{theorem}
	\label{thm:binaryUpg}
Let a BDMC $W:\calX \rightarrow \calY$ and an input distribution be given. 
Denote by $Q:\calX \rightarrow \calZ$ and $\Phi:\calZ \rightarrow \calY$ the optimizers of (\ref{eq:DIUpgStar}), for a given $L$. 
Denote by $\ppp{\bfy}_{y \in \calY}$ and $\ppp{\bfz}_{z \in \calZ}$ the posterior probabilities associated with the output letters of $W$ and $Q$, respectively.  
Assume without loss of generality that all the  $\{\bfz\}_{z \in \calZ}$ are distinct. 
Then, 
			\begin{equation}
			\label{eq:zSubsetOfY}
			\ppp{\bfz}_{z \in \calZ} \subseteq \ppp{\bfy}_{y \in \calY} \; .
			\end{equation}
Moreover, recalling the notation in (\ref{eq:yxzx}), $\Phi(y|z) > 0$ implies that $z$ either has the largest $z_0$ such that $z_0 \leq y_0$ or the smallest $z_0$ such that $z_0 \geq y_0$. 
\end{theorem}
The theorem essentially states that the optimal upgraded channel contains a subset of the output letters of the original channel, each such letter retaining its posterior probability vector.
Moreover, any output letter $y \in \calY$ is generated by the two output letters $z \in \calZ$ neighboring it, when ordered on the line which is the posterior probability simplex. 
Thus, if we are told the optimal subset $\ppp{\bfz}_{z \in \calZ}$  we can efficiently deduce from it the optimal $Q$ and $\Phi$. 
That is, to find $Q$, note that we can calculate the the probabilities $\ppp{\pi_z}$ using the RHS of (\ref{eq:piz}). 
After the above calculations are carried out, we have full knowledge of the reverse $Q$, and can hence deduce $Q$. 
From here, finding $\Phi$ is immediate: we can find the reverse $\Phi$ as in equation (\ref{eq:phiOptimal}), and then apply (\ref{eq:PhiReverse}) to get the forward channel.

Note that if $\bfz^{(a)}=\bfz^{(b)}$ for some two distinct letters $z^{(a)},z^{(b)} \in \calZ$, then we can merge these letters and obtain an equivalent channel~\cite[Section III]{Tal:Construct} with $L-1$ letters.
Repeating this until all the $\ppp{\bfz}_{z\in \calZ}$ are distinct allows us to assume distinction while retaining generality.

We stress that \Cref{thm:binaryUpg} is only valid for BDMCs, while for larger input alphabet sizes it can be disproved. 
For example, let $|\calX|=3$, and assume that the points $\{\bfy\}_{y \in \calY}$ are arranged on a circle in the simplex plane. 
Note now that no point can be expressed as a convex combination of the other points. 
Hence, (\ref{eq:bfy}) cannot be satisfied if $\{\bfz\}_{z \in \calZ}$ is a strict subset of $\{\bfy\}_{y \in \calY}$.
This example can be easily extended to larger input alphabet sizes using higher dimensional spheres.

\subsection{Optimal intermediate channel}
By definition, if $W \preccurlyeq Q$ then there exists a corresponding intermediate channel $\Phi$. 
Our aim now is to characterize the optimal $\Phi$ by which $\DIUpgStar$ in (\ref{eq:DIUpgStar}) is attained. 
Recall from (\ref{eq:bfy}) and (\ref{eq:PhiReverse}) our definition of the ``reverse'' channel $\Phi(z|y)=\Phi_{z|y}$. 
From (\ref{eq:DIUpgPhi}),
\begin{equation}
\label{eq:DIUpgSumIy}
\DIUpg = \sum_{y \in \calY} \pi_y h(\bfy) - \sum_{y \in \calY} \pi_y i_y \; ,
\end{equation}
where
\begin{equation}
\label{eq:iy}
i_y \triangleq \sum_{z \in \calZ} \Phi_{z|y}h(\bfz) \; .
\end{equation}
To recap, we have gained a simplification by considering reversed channels: each output letter $y \in \calY$ decreases $\DIUpg$ by $\pi_y i_y$. 

In the following lemma we consider a simple yet important case: an output letter $y$ of the original channel $W$ is gotten by combining exactly two output letters of the upgraded channel $Q$, denoted $z_1$ and $z_2$. 
Informally, the lemma states that the closer the posterior probabilities of $z_1$ and $z_2$ are to $y$, the better we are in terms of $i_y$.

\begin{lemma}
	\label{lm:iyMono}
	Let $\bfy=[\nonnegvar,1-\nonnegvar]^T$ be fixed. 
	For
\[
0 < \zeta_1 < \nonnegvar <\zeta_2 < 1 \; ,
\]
define $\bfz_1=[\zeta_1,1-\zeta_1]^T$ and $\bfz_2=[\zeta_2,1-\zeta_2]^T$. 
Next, let $\phi(\zeta_1,\zeta_2)$ be such that
	\begin{equation}
	\label{eq:iyConstraint}
	\bfy = \phi(\zeta_1,\zeta_2) \cdot \bfz_1 + (1-\phi(\zeta_1,\zeta_2)) \cdot \bfz_2 \; .
	\end{equation}
	Define
	\[
	i_y(\zeta_1,\zeta_2) \triangleq \phi(\zeta_1,\zeta_2)\cdot h(\bfz_1)+(1-\phi(\zeta_1,\zeta_2))\cdot h(\bfz_2) \; .
	\]
	Then, $i_y(\zeta_1,\zeta_2)$ is increasing with respect to $\zeta_1$ and decreasing with respect to $\zeta_2$.
\end{lemma}
\begin{IEEEproof}
	To satisfy (\ref{eq:iyConstraint}) we have
	\[
	\phi(\zeta_1,\zeta_2)= \frac{\zeta_2 - \nonnegvar}{\zeta_2-\zeta_1} \; .
	\]
	Now using the derivative of $i_y$ we get,
	\begin{align*}
	\frac{\pd i_y}{\pd \zeta_1} &= \frac{\pd \phi}{\pd \zeta_1} \cdot h(\bfz_1) + \phi \cdot \p{\eta'(\zeta_1)-\eta'(1-\zeta_1)} -\frac{\pd \phi}{\pd \zeta_1} \cdot h(\bfz_2) \\
	&=\frac{\pd \phi}{\pd \zeta_1}\cdot [ h(\bfz_1) + (\zeta_2-\zeta_1) (\eta'(\zeta_1)-\eta'(1-\zeta_1))\\
	&\quad -h(\bfz_2)] \\
	&= \frac{\pd \phi}{\pd \zeta_1} \cdot [ -h(\bfz_2) - \zeta_2 \log(\zeta_1)-(1-\zeta_2 )\log(1-\zeta_1) ] \\
	&=\frac{\zeta_2-\nonnegvar}{(\zeta_2-\zeta_1)^2} \dkl(\zeta_2||\zeta_1) \\
	&> 0 \; ,
	\end{align*}
	where we defined
	\begin{equation}
	\label{eq:KL}
	\dkl(p||q) \triangleq p \log \frac{p}{q} + (1-p) \log \frac{1-p}{1-q} \; ,
	\end{equation}
	which is the binary Kullback-Leibler divergence. 
	In the same manner,
	\begin{align*}
	\frac{\pd i_y}{\pd \zeta_2} &= \frac{\pd \phi}{\pd \zeta_2} \cdot h(\bfz_1) -\frac{\pd \phi}{\pd \zeta_2} \cdot h(\bfz_2)\\
	&\quad +(1-\phi)\cdot \p{\eta'(\zeta_2)-\eta'(1-\zeta_2)}\\
	&=\frac{\pd \phi}{\pd \zeta_2}\cdot [ h(\bfz_1)-h(\bfz_2) \\
	&\quad +(\zeta_2-\zeta_1)\p{\eta'(\zeta_2)-\eta'(1-\zeta_2)} ]\\
	&= \frac{\pd \phi}{\pd \zeta_2} \cdot [ h(\bfz_1) +\zeta_1 \log(\zeta_2) +(1-\zeta_1) \log (1-\zeta_2) ] \\
	&=-\frac{\nonnegvar-\zeta_1}{(\zeta_2-\zeta_1)^2} d(\zeta_1||\zeta_2) \\
	&< 0 \; ,
	\end{align*}
	and the proof is finished. 
\end{IEEEproof}

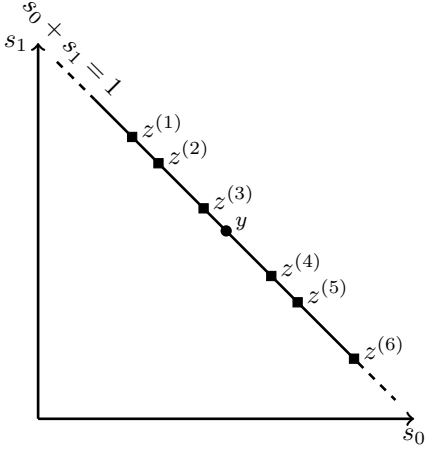
\begin{figure}[t]	
	\centering
	\begin{tikzpicture}[scale=5]
	\draw[->,line width=1pt] (0,0) -- (0,1)  node[left] {$s_1$};
	\draw[->,line width=1pt] (0,0) -- (1,0)  node[below] {$s_0$};
	\draw [line width=1pt] (0.15,0.85) -- (0.85,0.15) ;
	\draw[dashed,line width=1pt] (0.05,0.95) -- (0.15,0.85) node[pos=0.,sloped,above] {$s_0+s_1=1$} ;
	\draw[dashed,line width=1pt] (0.85,0.15) -- (0.95,0.05)   ;
	\fill[black] (0.5,0.5) node[right,yshift=0.1cm] {\footnotesize $y$} circle (0.015) ;
	\draw plot[mark=square*,mark size=0.35pt] coordinates{(0.25,0.75)} node[right,yshift=0.15cm]{$z^{(1)}$};	
	\draw plot[mark=square*,mark size=0.35pt] coordinates{(0.32,0.68)} node[right,yshift=0.15cm]{$z^{(2)}$};	
	\draw plot[mark=square*,mark size=0.35pt] coordinates{(0.44,0.56)} node[right,yshift=0.15cm]{$z^{(3)}$};	
	\draw plot[mark=square*,mark size=0.35pt] coordinates{(0.62,0.38)} node[right,yshift=0.15cm]{$z^{(4)}$};	
	\draw plot[mark=square*,mark size=0.35pt] coordinates{(0.69,0.31)} node[right,yshift=0.15cm]{$z^{(5)}$};	
	\draw plot[mark=square*,mark size=0.35pt] coordinates{(0.84,0.16)} node[right,yshift=0.15cm]{$z^{(6)}$};	
	\end{tikzpicture} 
	\caption{The letters $\ppp{z^{(i)}}_{i=1}^6$ of an upgraded BDMC $Q$ and a letter $y$ from an initial BDMC $W$. In this example $\calZ_L=\ppp{z^{(1)},z^{(2)},z^{(3)}}$, $\calZ_R=\ppp{z^{(4)},z^{(5)},z^{(6)}}$, $z_L^\ast=z^{(3)}$, $z_R^\ast=z^{(4)}$. }
	\label{fig:zypoints}
\end{figure}

The following lemma states that the second assertion of Theorem~\ref{thm:binaryUpg} holds.
\begin{lemma}
	\label{lm:optimalPhi}
Let a BDMC $W:\calX \rightarrow \calY$ and an input distribution be given. 
Denote by $Q:\calX \rightarrow \calZ$ and $\Phi:\calZ \rightarrow \calY$ the optimizers of (\ref{eq:DIUpgStar}), for a given $L$. 
Assume without loss of generality that all the $\{\bfz\}_{z \in \calZ}$ are distinct. 
Then, $\Phi(y|z) > 0$ implies that $z$ either has the largest $z_0$ such that $z_0 \leq y_0$ or the smallest $z_0$ such that $z_0 \geq y_0$.
\end{lemma}
\begin{IEEEproof}
	Note that the input distribution is given. 
	Thus, the reverse channels corresponding to $W$, $Q$, and $\Phi$ are well defined, and our proof will revolve around them.
	Since $W \preccurlyeq Q$, the reverse channel $\Phi_{z|y}$ satisfies (\ref{eq:bfy}), (\ref{eq:piz}) and (\ref{eq:DIUpgSumIy}).
	Let us assume to the contrary that $y\in \calY$ does not satisfy the assertion in the lemma.
	Our plan is to find a reverse channel $\Phi'_{z|y}$ that attains a greater $i_y$ than the one attained by $\Phi$, and thus arrive at a  contradiction.
	
	As a first case, assume that there exists $z^{\ast}\in \calZ$ for which $\bfz^{\ast}=\bfy$. 
	In this case, $z^\ast_0 = y_0$, and thus the lemma states that the only non-zero term in $\ppp{\Phi(y|z)}_{z \in \calZ}$ is  $\Phi(y|z^\ast)$. 
	Assume the contrary. 
	Thus, switching to the reverse channel, we note that $\ppp{\Phi_{z|y}}_{z \in \calZ}$ has at least two non-zero terms. 
	Then, using (\ref{eq:iy}) and the strict concavity of $h$ we get
	\[
	i_y < h\p{\sum_{z \in \calZ} \Phi_{z|y} \bfz  } = h(\bfy) \; .
	\] 
	Note that this upper bound can be attained by simply choosing $\Phi'_{z|y}=1$ when $z=z^{\ast}$ and $\Phi'_{z|y}=0$ otherwise.
	For all other $y' \neq y$, we define $\Phi'_{z|y'}$ as equal to $\Phi_{z|y'}$.
	Thus, using $\Phi'_{z|y}$ and (\ref{eq:piz}), we obtain a new set of probabilities $\ppp{\pi'_z}_{z \in \calZ}$ that together with the reverse channel $Q$ generate a new ``forward'' channel $Q' : \calX \to \calZ$ (applying Bayes' rule).
	Since the reverse channels satisfy $Q' \succcurlyeq W$, we have by (\ref{eq:bfy}) and the explanation following it that the ``forward'' channels satisfy the same upgrading relation. 
	Yet $Q'$ attains a strictly lower $\DIUpg$, a contradiction.
	
	In the second case, we assume that $\bfz \neq \bfy$ for all $z \in \calZ$ and define the sets
	\begin{align*}
		\calZ_R = \ppp{z \in \calZ: z_0>y_0 } \; ,\\
		\calZ_L = \ppp{z \in \calZ: z_0<y_0 } \; .
	\end{align*}
	Geometrically, if we draw $\bfy$ and $\ppp{\bfz}_{z \in \calZ}$ as points on the line $s_0+s_1=1$ in the first quadrant of the $(s_0,s_1)$ plane (the two dimensional simplex), then the letters in $\calZ_R$ would be on the right side of $\bfy$ and the letters in $\calZ_L$ on its left (see \Cref{fig:zypoints}). 
	Define also
	\[
	z_L^{\ast} \triangleq \argmin_{z \in \calZ_L} \norm{\bfy-\bfz}_2 \; ,  \quad z_R^{\ast} \triangleq \argmin_{z \in \calZ_R} \norm{\bfy-\bfz}_2 \; .
	\]
	Namely, the closest neighboring letters from $\calZ$, one from each side.
	The lemma states that the only non-zero terms in $\ppp{\Phi(y|z)}_{z \in \calZ}$ are  $\Phi(y|z_L^\ast)$ and $\Phi(y|z_R^\ast)$.
	Assume again the contrary.
	Thus, switching again to the reverse channel, both $\ppp{\Phi_{z|y}}_{z \in \calZ_L}$ and $\ppp{\Phi_{z|y}}_{z \in \calZ_R}$ have non-zero terms.
	By assumption, one such term corresponds to neither $z_L^\ast$ nor $z_R^\ast$.
	Note that we can write
	\begin{align*}
	i_y &= \sum_{z\in \calZ_L} \Phi_{z|y} h(\bfz) + \sum_{z\in \calZ_R} \Phi_{z|y} h(\bfz) \\
	&=\phi \sum_{z\in \calZ_L} \frac{\Phi_{z|y}}{\phi} h(\bfz) + (1-\phi) \sum_{z\in \calZ_R} \frac{\Phi_{z|y}}{1-\phi} h(\bfz) \; ,
	\end{align*}
	where $\phi \triangleq \sum_{z \in \calZ_L} \Phi_{z|y}$, and $0 < \phi < 1$. 
	Using the strict concavity of $h$ we get
	\begin{align}
	\label{eq:iyMax}	
	i_y & \leq \phi \cdot h\p{\sum_{z\in \calZ_L} \frac{\Phi_{z|y}}{\phi} \bfz} + (1-\phi) \cdot h\p{\sum_{z\in \calZ_R} \frac{\Phi_{z|y}}{1-\phi} \bfz} \nonumber \\
	&= \phi \cdot h(\bfz_L) + (1-\phi) \cdot h(\bfz_R)     \; , 	
	\end{align}
	where
	\[
	\bfz_L \triangleq \sum_{z\in \calZ_L} \frac{\Phi_{z|y}}{\phi} \bfz \; , \quad \bfz_R \triangleq \sum_{z\in \calZ_R} \frac{\Phi_{z|y}}{1-\phi} \bfz \; .
	\]
	Note that the locations of $\bfz_L$ and $\bfz_R$ depend on the probabilities $\ppp{\Phi_{z|y}}_{z \in \calZ}$.
	Since both $\bfz_L$ and $\bfz_R$ are convex combinations of the letters in $\calZ_L,\calZ_R$, respectively, they have to reside in the convex hull of those sets. 
	Then, by assumption, either $\bfz_L \neq \bfz_L^\ast$ or $\bfz_R \neq \bfz_R^\ast$.
	Recall now that according to \Cref{lm:iyMono}, any choice of $\ppp{\Phi_{z|y}}_{z \in \calZ}$ could be improved as long as $\bfz_L$ and $\bfz_R$ are not the closest letters to $\bfy$.
	Hence,
	\[
	i_y< \phi' \cdot h(\bfz_L^\ast) + (1-\phi') \cdot h(\bfz_R^\ast)     \; ,
	\]
	for the corresponding $\phi'$.
	Once again, this upper bound can be attained by choosing
	\begin{equation}
	\label{eq:phiOptimal}
	\Phi'_{z|y}=
	\begin{cases}
	\frac{\|\bfz_R^{\ast}-\bfy\|_2}{\|\bfz_R^{\ast}-\bfz_L^{\ast}\|_2} & z=z_L^{\ast} \; , \\[0.2cm]
	\frac{\|\bfy-\bfz_L^{\ast}\|_2}{\|\bfz_R^{\ast}-\bfz_L^{\ast}\|_2} & z=z_R^{\ast} \; , \\[0.2cm]
	0 & \mbox{Otherwise}\; .
	\end{cases}
	\end{equation}
	Thus, as before, we have found a channel $Q' \succcurlyeq W$ that attains a strictly lower $\DIUpg$, a contradiction.
\end{IEEEproof}

\subsection{Optimal upgraded channel}
\Cref{lm:optimalPhi} is meaningful for two reasons. 
First, now that we know the optimal $\Phi$ for any given $Q$, we can minimize over $Q$ alone. 
Equivalently, as per the discussion after Theorem~\ref{thm:binaryUpg}, we can minimize over the subset $\ppp{\bfz}_{z \in \calZ}$. 
Second, any two adjacent letters in $\ppp{\bfz}_{z \in \calZ}$  (on the simplex line) exclusively generate all the letters $\ppp{\bfy}$ that reside in the segment between the two of them. 
Hence, our proof of \Cref{thm:binaryUpg} can be ``localized" in the sense that we can focus on two adjacent segments and move the separating letter $\bfz$ between them, thus only affecting the letters $\ppp{\bfy}$ in those two segments.
We return to the proof of our theorem.

\begin{IEEEproof}[Proof of \Cref{thm:binaryUpg}]
	Since Lemma~\ref{lm:optimalPhi} has been proved, all that is left is to prove the inclusion (\ref{eq:zSubsetOfY}).
	Note that using (\ref{eq:bfy}) and $\sum_{z \in \calZ} \Phi_{z|y}=1$, if $W \preccurlyeq Q$ then
	\[
	\ppp{\bfy}_{y \in \calY} \subseteq \conv \ppp{\ppp{\bfz}_{z \in \calZ}} \; ,
	\]
	where $\mathrm{conv}$ denotes convex hull. 
	Namely, each $\bfy$ can be expressed as a convex combination of vectors $\ppp{\bfz}$ that correspond to letters in $\calZ$.
	
	Let us assume first that $L=2$. 
	That is, we need to find two letters $\bfz_{\min}$ and $\bfz_{\max}$ whose convex hull contains all the letters $\ppp{\bfy}_{y \in \calY}$ and $\DIUpg$ is minimized.
	Then, according to \Cref{lm:iyMono}, it is optimal to choose $\bfz_{\min}$ and $\bfz_{\max}$ with the smallest possible convex hull containing $\ppp{\bfy}_{y \in \calY}$.
	Hence, the letters of the optimal $Q$ are
	\begin{equation}
	\label{eq:yLyR}
	\bfz_{\min}=\bfy_{\min} \; , \quad  \bfz_{\max}=\bfy_{\max} \; ,
	\end{equation}
	where
	\[
	y_{\min} \triangleq \argmin_{y \in \calY} y_0 \; , \quad y_{\max} \triangleq \argmax_{y \in \calY} y_0 \; .
	\]
	Namely, the leftmost and rightmost letters in $\ppp{\bfy}_{y \in \calY}$, respectively.
	
	Assume now that $L>2$ and let 
	\[
	\bfz^{(1)}=[\zeta_1,1-\zeta_1]^T \; , \quad \bfz=[\zeta,1-\zeta]^T \; , \quad \bfz^{(2)}=[\zeta_2,1-\zeta_2]^T \; ,
	\]
	be contiguous points on the simplex line satisfying
\[
0 \leq \zeta_1 < \zeta < \zeta_2 \leq 1 \; . 
\]
	Assume there is a subset $\calYTilde \subseteq \calY$ of $M$ letters in the \emph{interior} of $\conv \ppp{\bfz^{(1)},\bfz^{(2)}}$. 
	Thus, we stress that $z^{(1)}$ and $z^{(2)}$ are not contained in $\calYTilde$.
	Our aim is to show that an optimal choice for $\bfz$ satisfies $\bfz \in \calYTilde$. 
	That will ensure that there cannot be a letter $z \in \calZ$ such that $\bfz \notin \ppp{\bfy}_{y \in \calY}$, except maybe for the two extreme ones (since $\bfz$ is internal). 
	However, as in the $L=2$ case discussed previously, these two extreme letters must be $\bfy_{\min}$ and $\bfy_{\max}$, as defined above.
	
	Note that if $M=0$ then $\pi_z=0$, by \Cref{lm:optimalPhi}. 
	In this case, without loss of generality, $\bfz$ can be removed from $\calZ$. 
	Thus, we henceforth assume that $M>0$. 
	\Cref{fig:zypoints2} illustrates some of the sets and letters we will shortly define. 
	
	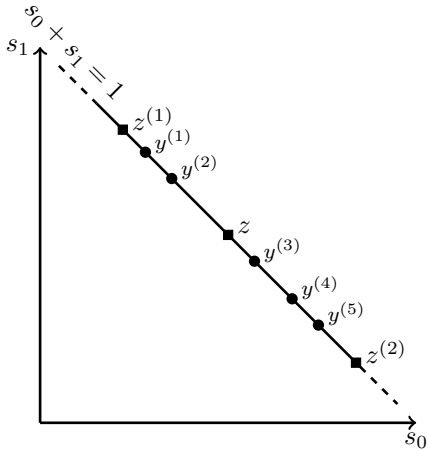
\begin{figure}[t]	
		\centering
		\begin{tikzpicture}[scale=5]
		\draw[->,line width=1pt] (0,0) -- (0,1)  node[left] {$s_1$};
		\draw[->,line width=1pt] (0,0) -- (1,0)  node[below] {$s_0$};
		\draw [line width=1pt] (0.15,0.85) -- (0.85,0.15) ;
		\draw[dashed,line width=1pt] (0.05,0.95) -- (0.15,0.85) node[pos=0.,sloped,above] {$s_0+s_1=1$} ;
		\draw[dashed,line width=1pt] (0.85,0.15) -- (0.95,0.05)   ;
		\fill[black] (0.28,0.72) node[right,yshift=0.15cm] {\footnotesize $y^{(1)}$} circle (0.015) ;
		\fill[black] (0.35,0.65) node[right,yshift=0.15cm] {\footnotesize $y^{(2)}$} circle (0.015) ;
		\fill[black] (0.57,0.43) node[right,yshift=0.15cm] {\footnotesize $y^{(3)}$} circle (0.015) ;		
		\fill[black] (0.67,0.33) node[right,yshift=0.15cm] {\footnotesize $y^{(4)}$} circle (0.015) ;		
		\fill[black] (0.74,0.26) node[right,yshift=0.15cm] {\footnotesize $y^{(5)}$} circle (0.015) ;		
		\draw plot[mark=square*,mark size=0.35pt] coordinates{(0.22,0.78)} node[right,yshift=0.15cm]{$z^{(1)}$};	
		\draw plot[mark=square*,mark size=0.35pt] coordinates{(0.5,0.5)} node[right,yshift=0.1cm]{$z$};	
		\draw plot[mark=square*,mark size=0.35pt] coordinates{(0.84,0.16)} node[right,yshift=0.15cm]{$z^{(2)}$};	
		\end{tikzpicture} 
		\caption{The letters $z^{(1)},z$ and $z^{(2)}$ of an upgraded BDMC $Q$, and $\ppp{y^{(i)}}_{i=1}^5$ of an initial BDMC $W$. 
			In this example, when $z$ is picked as illustrated: $\calYTilde=\ppp{y^{(i)}}_{i=1}^5$, $\calYTilde_L=\ppp{y^{(1)},y^{(2)}}$, $\calYTilde_R=\ppp{y^{(3)},y^{(4)},y^{(5)}}$, $\protect \yutilde =y^{(2)}$, $\ytilde=y^{(3)}$.       }
		\label{fig:zypoints2}
	\end{figure}
	
	If $\bfz \in \ppp{\bfy}_{y \in \calY}$, then we are done. 
	Henceforth, let us assume this is not the case. 
	We express each $\bfy$ as $\bfy=[y_0 ,1-y_0]^T$, and partition $\calYTilde$ into the sets
	\[
	\calYTilde_L \triangleq \ppp{y \in \calYTilde : y_0 < \zeta} \; , \quad \calYTilde_R=\ppp{y \in \calYTilde:y_0>\zeta} \; ,
	\]
	for a given $\zeta$. 
	Since $M>0$, at least one subset is non-empty. 
	In fact, by similar reasoning to the $L=2$ case, we deduce that both $\calYTilde_L$ and $\calYTilde_R$ are non-empty. 
	By (\ref{eq:iy}) and \Cref{lm:optimalPhi}, the contribution of these subsets to $\DIUpg$ is
	\begin{align*}
	\DIUpg(\calYTilde) &\triangleq  \sum_{y \in \calYTilde} \pi_y h(\bfy) - \sum_{y \in \calYTilde_L} \pi_y i_y- \sum_{y \in \calYTilde_R} \pi_y i_y   \\
	&= \sum_{y \in \calYTilde} \pi_y h(\bfy)  \nonumber \\
	&\quad  - \sum_{y \in \calYTilde_L} \pi_y \p{ \frac{\zeta-y_0}{\zeta-\zeta_1} h(\bfz^{(1)}) + \frac{y_0-\zeta_1}{\zeta-\zeta_1} h(\bfz) }   \nonumber  \\
	&\quad - \sum_{y \in \calYTilde_R} \pi_y \p{ \frac{\zeta_2-y_0}{\zeta_2-\zeta} h(\bfz) + \frac{y_0-\zeta}{\zeta_2-\zeta}  h(\bfz^{(2)})        } \; .  \nonumber
	\end{align*}
	Thus,
	\begin{multline}
	\label{eq:DIUpgYTilde}
	\DIUpg(\calYTilde) =\sum_{y \in \calYTilde} \pi_y \p{h(\bfy)-h(\bfz^{(1)})-h(\bfz^{(2)})}     \\
	 -C_1 \frac{h(\bfz)-h(\bfz^{(1)})}{\zeta-\zeta_1} +C_2 \frac{h(\bfz^{(2)})-h(\bfz)}{\zeta_2-\zeta} \; ,
	\end{multline}
	where
	\[
	C_1 \triangleq \sum_{y \in \calYTilde_L} \pi_y(y_0-\zeta_1) \; , \quad C_2 \triangleq \sum_{y \in \calYTilde_R} \pi_y(\zeta_2-y_0) \; .
	\]
	Recall that $\calYTilde_L,\calYTilde_R$ were defined as a function of $\bfz$. 
	Hence, they remain the same as long as $\bfz$ is strictly between the rightmost letter in $\calYTilde_L$ and the leftmost letter in $\calYTilde_R$, denoted by $\bfyutilde \triangleq [\yutilde_0,1-\yutilde_0]^T$ and $\bfytilde \triangleq [\ytilde_0,1-\ytilde_0]^T$, respectively. 

By definition, $\calYTilde$ does not contain $z^{(1)}$ and $z^{(2)}$. This implies the following two points. 
First, we deduce that $C_1$ and $C_2$ are both positive. 
Next,  by (\ref{eq:DIUpgYTilde}), $\DIUpg(\calYTilde)$ is continuous and bounded for $\yutilde_0 \leq \zeta \leq \ytilde_0$. 
The theorem will be proved if we show that the minimum is attained by setting $\zeta$ to either $\yutilde_0$ or $\ytilde_0$. 
Thus, we show that the minimum is not attained when $\zeta \in (\yutilde_0,\ytilde_0)$.
To that end, we take the derivative of $\DIUpg(\calYTilde)$ with respect to $\zeta$ and get
	\begin{align*}
		\frac{\pd \DIUpg(\calYTilde)}{\pd \zeta} &= -C_1\frac{(\eta'(\zeta)-\eta'(1-\zeta) ) (\zeta-\zeta_1)   }{(\zeta-\zeta_1)^2} \\
		&\quad -C_2 \frac{(\eta'(\zeta)-\eta'(1-\zeta))(\zeta_2-\zeta)     }{(\zeta_2-\zeta)^2} \\
		& \quad +C_1\frac{h(\bfz)-h(\bfz^{(1)})   }{(\zeta-\zeta_1)^2} + C_2 \frac{h(\bfz^{(2)})-h(\bfz)}{(\zeta_2-\zeta)^2} \\
		&= C_1 \frac{\dkl(\zeta_1||\zeta)}{(\zeta-\zeta_1)^2}-C_2 \frac{\dkl(\zeta_2||\zeta)}{(\zeta_2-\zeta)^2} \; .
	\end{align*}
	We now recall that both $C_1$ and $C_2$ are positive. 
	Thus,
	\[
	\frac{\pd \DIUpg(\calYTilde)}{\pd \zeta} = C_2 \frac{\dkl(\zeta_1||\zeta)}{(\zeta-\zeta_1)^2} \p{\frac{C_1}{C_2}- q(\zeta) } \; ,
	\]
	where we defined
	\[
	q(\zeta) \triangleq \frac{\dkl(\zeta_2||\zeta) (\zeta-\zeta_1)^2  }{\dkl(\zeta_1||\zeta)(\zeta_2-\zeta)^2} \; .
	\]
	To achieve our goal, it suffices to show that $q(\zeta)$ is non-decreasing in $(\zeta_1,\zeta_2)$, thus ensuring that $\DIUpg(\calYTilde)$ is either monotonic or has a single maximum. 
	The proof is given in Appendix \ref{app:OBU}.
\end{IEEEproof}

\subsection{Dynamic programming implementation}
\Cref{thm:binaryUpg} simplifies the task of channel upgrading to finding the optimal $L$-sized subset of $\calY$.
Note that such a subset must contain the leftmost and rightmost letters of $\calY$.

We can now use dynamic programming to efficiently find the optimal subset. 
The key idea is to use the structure of~(\ref{eq:DIUpgSumIy}) and the ordering of the letters on the simplex.
Each possible $L$-sized subset partitions $\calY$ to $L-1$ contiguous sets on the simplex, with overlapping borders (the internal $\ppp{\bfz}$).
Since the cost function (\ref{eq:DIUpgSumIy}) is additive in the letters $\calY$, we can employ a dynamic programming approach, similar to~\cite[Section IV]{Kurkoski:Quantization} and \cite[Section III]{Iwata:SMAWK}.

\section{Upper bound on binary optimal upgrading gain}
\label{sec:UBBUC}
Our final result is the fruit of combining \Cref{sec:UBDC}, \Cref{sec:LBUC} and \Cref{sec:OBU}. 
Namely, an upper bound on $\DIUpgStar$ and $\UC$ for $|\calX|=2$. 
Once again, we use an iterative sub-optimal upgrading algorithm called ``greedy-split'', similar to the one proposed in~\cite{Pedarsani:Construction}. 
Implicitly, we apply the optimal upgrading algorithm, to get from an alphabet of $|\calY|$ output letters to one with $|\calY|-1$ output letters. 
This is done iteratively, until the required number of letters, $L$, is reached. 
This simplifies to the following. 
In each iteration we find the letter $y \in \calY$ that minimizes
\begin{equation}
\label{eq:DIUpgBinary}
\DIUpg = \pi_y h(\bfy)-\pi_y \phi h(\bfyleft) -\pi_y (1-\phi) h(\bfyright) \; ,
\end{equation}
where $\yleft$ and $\yright$ are the left and right adjacent letters to $y$, respectively, and
\[
\phi = \frac{\norm{\bfyright-\bfy}_2}{\norm{\bfyright-\bfyleft}_2} \; ,
\]
 as in  \Cref{lm:optimalPhi}. 
 The minimizing letter $y$ is then split between $\yleft$ and $\yright$ by updating
 \[
 \pi_{\yleft} \leftarrow \pi_{\yleft} +\phi \cdot \pi_y \; , \quad \pi_{\yright} \leftarrow \pi_{\yright} + (1-\phi) \cdot \pi_{y} \; ,
 \]
 and then eliminating $y$. 
 The following theorem is the main result of this section.
 \begin{theorem}
 	\label{thm:DIUpgStar}
 	Let a BDMC $W:\calX \rightarrow \calY$ satisfy $|\calY|>8$, and let $L \geq 8$. 
 	Then, for any fixed input distribution,
 	\[
 	\DIUpgStar =\min_{\substack{{Q,\Phi:Q \succcurlyeq W,} \\ {|Q| \leq L}} } I(Q)-I(W)= O\p{L^{-2}} \; .
 	\] 
 	In particular,
 	\[
 	\DIUpgStar \leq 2\nu(2)\cdot L^{-2} \; ,
 	\]
 	where $\nu(\cdot)$ was defined in \Cref{thm:DIDegStar}. 
 	This bound is attained by greedy-split and is tight in the power-law sense.
 \end{theorem}

As in \Cref{sec:UBDC}, we first prove the following theorem.
\begin{theorem}
	\label{thm:DIUpgBinary} 
	Let a BDMC $W:\calX \rightarrow \calY$ satisfy $|\calY|>8$, and let a BDMC $Q:\calX \rightarrow \calZ$ be the result of upgrading $W$ by splitting a letter $y \in \calY$. 
	Then, for any fixed input distribution, there exists a letter $y \in \calY$ for which the resulting $Q$ satisfies
	\[
	\DIUpg = O\p{|\calY|^{-3}} \; .
	\]
	In particular,
	\begin{equation}
	\label{eq:DIUpgFullBound}
	\DIUpg \leq   2\mu(2)\cdot |\calY|^{-3} \; ,
	\end{equation}
	where $\mu(\cdot)$ was defined in \Cref{thm:DIDeg}. 
\end{theorem}
\begin{IEEEproof}
	Note that (\ref{eq:DIUpgBinary}) has the form of (\ref{eq:DIxsum}).
	That is, $\bfy_L$ plays the role of $\bfalpha$; $\bfy_R$ plays the role of $\bfbeta$; $\pi_y\phi$ plays the role of $\pi_{\alpha}$; $\pi_y(1-\phi)$ plays the role of $\pi_{\beta}$; $\bfy$ plays the role of $\bfgamma$; the first $\pi_y$ in (\ref{eq:DIUpgBinary}) plays the role of $\pi_{\gamma}$.
	
	Thus, (\ref{eq:DIBound}) applies to our case as well, yielding
	\begin{equation}
	\label{eq:DIUpgBinaryBound}
	\DIUpg \leq  2\pi_y \cdot d(\bfyleft,\bfyright) \; ,
	\end{equation}
	where $d$ was defined in (\ref{eq:dDef}).
	As we did before, we narrow the search to letters in $\calYSmall$. 
	Note that $\yleft$ and $\yright$ cannot be adjacent, hence we define $\calYPunctured$ as the subset of $\calYSmall$ one gets by eliminating every other letter, starting from the second letter, when drawing $\ppp{\bfy}_{y\in\calYSmall}$ on the two dimensional simplex. 
	Thus, each pair of adjacent letters in $\calYPunctured$ has a letter from $\calYSmall$ in its convex hull. 
	This operation of puncturing results in
	\begin{equation}
	\label{eq:YpuncturedSize}
	|\calYPunctured| \geq \frac{|\calYSmall|}{2} \geq \frac{|\calY|}{4} \; .
	\end{equation}
	Using the same method as in \Cref{thm:DIDeg}, there exists a pair $\yleft,\yright \in \calYPunctured$ for which
	\begin{equation}
	\label{eq:BinaryUpgClosePair}
	d(\bfyleft,\bfyright) \leq 4 \cdot r \; ,
	\end{equation}
	where $r$ was defined in (\ref{eq:rstarDef}). 
	The factor of $4$ is due to the following.
	Since we are using $\calYPunctured$, the proof of Theorem~\ref{thm:DIDeg} must be slightly changed towards the end. 
	Specifically,~(\ref{eq:calYTagSize}) changes to
	\[
	|\calY'| \geq \frac{|\calYPunctured|}{|\calX|}\geq \frac{|\calYSmall|}{2|\calX|}  \geq \frac{|\calY|}{4|\calX|} > 1 \; .
	\]
	Then, the term $|\calY|/(2|\calX|)$ in~(\ref{eq:spherePackingDerivation}) becomes $|\calY|/(4|\calX|)$. 
	Thus, the $r$ needed for the starred equality in (\ref{eq:spherePackingDerivation}) to hold is $4$ times the original $r$.
	Now, plugging (\ref{eq:BinaryUpgClosePair}) in (\ref{eq:DIUpgBinaryBound}) and recalling that $\pi_y \leq 2/|\calY|$ for $y\in \calYSmall$, we get (\ref{eq:DIUpgFullBound}).
\end{IEEEproof}
\begin{IEEEproof}[Proof of \Cref{thm:DIUpgStar}]
	The proof uses \Cref{thm:DIUpgBinary} iteratively, and is similar to \Cref{thm:DIDegStar}, hence omitted. 
	The tightness in power-law is due to \Cref{thm:DIUpg}.	
\end{IEEEproof}

Note that the greedy-split algorithm can be implemented in a similar manner to~\cite[Algorithm C]{Tal:Construct}.
Hence, it has the same complexity, that is, $O(|\calY| \log |\calY|)$.

We now discuss the application of \Cref{thm:DIUpgStar} to symmetric channels, as we did in \Cref{subsec:symmetricChannels}.
Assume that $W$ is a symmetric channel according to our definition in \Cref{subsec:symmetricChannels}.
To upgrade $W$ to a symmetric channel $Q$, we slightly modify the greedy-split algorithm as follows.
Instead of searching over all $y \in \calY$, we limit the search to all $y \in \calY$ for which $W(0|y)<\frac{1}{2}$ and $W(0|y_R) \leq \frac{1}{2}$.
Then, after splitting $y$ to $y_L$ and $y_R$, we split $\bar{y}$ to $\bar{y}_L$ and $\bar{y}_R$ where $\bar{y}$, $\bar{y}_L$ and $\bar{y}_R$ are the corresponding ``conjugate'' letters in the partition of $\calY$, as defined in \cite[Section II]{Tal:Construct}.
Using the same arguments as in the proof of \Cref{theo:symmetric}, we deduce that \Cref{thm:DIUpgBinary} holds for the symmetric case as well.

\iftoggle{IEEEtran}{\appendices}{\appendix}
\section{Last claim in the proof of Theorem~\ref{thm:binaryUpg}}
\label{app:OBU}

Recall that the proof of Theorem~\ref{thm:binaryUpg} will be complete, once the last step in it is justified. 
Thus, we state and prove the following lemma.
\begin{lemma}
\label{lemm:proofAddendum}
Fix $0 \leq \zeta_1 < \zeta_2 \leq 1$. For $\zeta_1 < \zeta < \zeta_2$, define
\[
q(\zeta) \triangleq \frac{\dkl(\zeta_2||\zeta) (\zeta-\zeta_1)^2  }{\dkl(\zeta_1||\zeta)(\zeta_2-\zeta)^2} \; .
\]
Then, $q(\zeta)$ is non-decreasing.
\end{lemma}

\begin{IEEEproof}
To prove that $q(\zeta)$ is non-decreasing, we consider its derivative. 
Straightforward algebraic manipulations yield that
\begin{multline}
	\label{eq:q2Def}
	\frac{\dd q}{\dd \zeta} = \frac{2(\zeta-\zeta_1)(\zeta_2-\zeta)\dkl(\zeta_1||\zeta)\dkl(\zeta_2||\zeta)(\zeta_2-\zeta_1)}{\pp{\dkl(\zeta_1||\zeta)(\zeta-\zeta_2)^2}^2}  \\
	 \quad \cdot \bigg[ 1- \frac{(\zeta_2-\zeta)^2}{2\zeta(1-\zeta)\dkl(\zeta_2||\zeta)}  \cdot  \frac{\zeta-\zeta_1}{\zeta_2-\zeta_1} -  \\
	\quad \quad \frac{(\zeta_1-\zeta)^2}{2\zeta(1-\zeta)\dkl(\zeta_1||\zeta)} \cdot   \frac{\zeta_2-\zeta}{\zeta_2-\zeta_1}   \bigg] \; .
\end{multline}
Note that the term outside the brackets is non-negative. 
For the inner term, let us define the two-variable function
\[
q_2(\tau,\zeta) \triangleq 
\begin{cases}
\frac{(\tau-\zeta)^2}{2\zeta(1-\zeta)\dkl(\tau||\zeta)} & \zeta \neq \tau \; , \\[0.2cm]
1 & \zeta=\tau \; .
\end{cases}
\]
For $0 < \zeta < 1$ fixed, $q_2(\tau,\zeta)$ is a continuous function of $\tau$, where $0 < \tau < 1$.
Indeed, this can be deduced by a double application of L'H\^opital's rule.
As we will show, more is true: for $\zeta$ and $\tau$ as above, $q_2(\tau,\zeta)$ is concave as a function of $\tau$.
Namely, for every $\tau_1,\tau_2,\theta \in [0,1]$,
\[
q_2(\theta \tau_1 +(1-\theta) \tau_2,\zeta) - \theta \cdot q_2 (\tau_1,\zeta) - (1-\theta)\cdot q_2 (\tau_2,\zeta) \geq 0 \; .
\]
By choosing $\tau_1=\zeta_2$, $\tau_2=\zeta_1$, and $\theta=\frac{\zeta-\zeta_1}{\zeta_2-\zeta_1}$, we get the non-negativity of the inner term in (\ref{eq:q2Def}).

To prove the  concavity of $q_2(\tau,\zeta)$ with respect to $\tau$, we will show that the second derivative $\frac{\pd^2 q_2}{\pd \tau^2}$ is non-positive.
The following identity is easily proved, and will be used to simplify many otherwise unwieldy expressions:
\begin{equation}
\label{eq:KLSimplification}
\boxed{
(p-q) \frac{\pd \dkl(p||q)}{\pd p} = \dkl(p||q) + \dkl(q||p) \; .
}
\end{equation}
Using the above, we deduce that 
\[
\frac{\pd q_2}{\pd \tau} =
\begin{cases}
\frac{(\tau-\zeta)\left[ \dkl(\tau||\zeta)-\dkl(\zeta||\tau) \right]}{2 \zeta(1-\zeta)(\dkl(\tau||\zeta))^2}  & \zeta \neq \tau \; , \\
\frac{1- 2\zeta}{3\zeta(1-\zeta)} & \zeta = \tau \; ,
\end{cases}
\]
where the derivative for the case $\zeta = \tau$ is obtained by considering the limit
\[
\lim_{\tau \to \zeta} \frac{q_2(\tau,\zeta) - q_2(\zeta,\zeta)}{\tau - \zeta} \; .
\]
By an application of L'H\^opital's rule, we deduce that $\frac{\pd q_2}{\pd \tau}$ is continuous in $\tau$.

We differentiate once again, using (\ref{eq:KLSimplification}), simple algebra, and similar reasoning to what was employed before. 
The result is
\[
\frac{\pd^2 q_2}{\pd \tau^2}=
\begin{cases}
\frac{(\tau-\zeta)^2}{2\zeta(1-\zeta)\dkl(\tau||\zeta)^3 \tau(1-\tau)} & \\
\quad \cdot \p{\frac{2\tau(1-\tau)\dkl(\zeta||\tau)^2}{(\tau-\zeta)^2}-\dkl(\tau||\zeta)} & \zeta \neq \tau \; , \\
\frac{\zeta - \zeta^2 -1}{9(1-\zeta)^2\zeta^2} & \zeta = \tau \; ,
\end{cases}
\]
a continuous function.

Clearly, $\frac{\pd^2 q_2}{\pd \tau^2}$ is negative when $\zeta = \tau$, since $0 < \zeta < 1$. 
Considering the case $\zeta \neq \tau$, we see that $\frac{\pd^2 q_2}{\pd \tau^2}$ is non-positive when the bracketed term is non-positive,
\[
\frac{2\tau(1-\tau)\dkl(\zeta||\tau)^2}{(\tau-\zeta)^2} \leq \dkl(\tau||\zeta) \; .
\]
Note now that both sides go to zero as $\zeta \to \tau$. 
The derivatives of the LHS and RHS with respect to $\zeta$ are
\[
\frac{4 \tau (1-\tau) \dkl (\zeta || \tau) \dkl (\tau || \zeta)}{( \zeta - \tau)^3}
\]
and
\[
\frac{\zeta-\tau}{\zeta(1-\zeta)} \; , 
\]
respectively. 
Both derivatives are positive when $\zeta > \tau$, negative when $\zeta < \tau$ and zero when $\zeta=\tau$. 
Thus it suffices to show that the ratio between the derivatives (left over right) is less than $1$. 
That is,
\[
\frac{2\zeta (1-\zeta)\dkl(\tau||\zeta)}{(\zeta-\tau)^2}\cdot \frac{2\tau(1-\tau)\dkl(\zeta||\tau)}{(\tau-\zeta)^2} \leq 1 \; ,
\]
which is equivalent when taking the square root of both sides. 
Using the AM-GM inequality it is enough to show that
\[
\frac{1}{2} \frac{2\zeta (1-\zeta)\dkl(\tau||\zeta)}{(\zeta-\tau)^2} + \frac{1}{2} \frac{2\tau(1-\tau)\dkl(\zeta||\tau)}{(\tau-\zeta)^2} \leq 1 \; ,
\]
which is equivalent to showing
\begin{align*}
q_3(\tau,\zeta) &\triangleq \zeta(1-\zeta) \dkl(\tau||\zeta) + \tau(1-\tau) \dkl(\zeta||\tau) - (\tau-\zeta)^2\\
& \leq 0 \; .
\end{align*}
The function $q_3$ can be shown to be twice continuously differentiable with respect to $\tau$ for a fixed $\zeta \in (0,1)$. 
Its second derivative satisfies
\[
\frac{\pd^2 q_3}{\pd \tau^2}(\tau,\zeta) =   - \frac{(\zeta-\tau)^2+2\tau(1-\tau)\dkl(\zeta||\tau)}{\tau(1-\tau)} \leq 0 \; ,
\]
and thus $q_3$ is concave with respect to $\tau$. 
Moreover,
\[
\frac{\pd q_3}{\pd \tau}(\zeta,\zeta) = 0 \; ,
\]
which means that $q_3$ is maximal when $\tau=\zeta$, namely,
\[
q_3(\tau,\zeta) \leq q_3(\zeta,\zeta)=0 \; ,
\]
and the proof is finished.
\end{IEEEproof}

\section*{Acknowledgements}
We thank Boaz Shuval for the helpful suggestions.

\end{document}

%% file: DegradingUpgrading_Shortcuts.tex
\newcommand {\calX} {\mathcal{X}}
\newcommand {\calY} {\mathcal{Y}}
\newcommand {\calZ} {\mathcal{Z}}
\newcommand {\calB} {\mathcal{B}}
\newcommand {\calC} {\mathcal{C}}
\newcommand {\calQ} {\mathcal{Q}}
\newcommand {\calV} {\mathcal{V}}
\newcommand {\calU} {\mathcal{U}}
\newcommand {\calS} {\mathcal{S}}

\DeclareMathOperator*{\argmin}{arg\,min}
\DeclareMathOperator*{\argmax}{arg\,max}
\newcommand{\p}[1]{\left(#1\right)}
\newcommand{\pp}[1]{\left[#1\right]}
\newcommand{\ppp}[1]{\left\{#1\right\}}

\newcommand{\DIDeg}{\Delta I_{\shortdownarrow}}
\newcommand{\DIUpg}{\Delta I_{\shortuparrow}}
\newcommand{\DIDegStar}{\Delta I_{\shortdownarrow}^{\ast}}
\newcommand{\DIUpgStar}{\Delta I_{\shortuparrow}^{\ast}}

\newcommand {\DC} {\mathrm{DC}}
\newcommand {\UC} {\mathrm{UC}}

\newcommand{\nonnegvar}{\sigma}

\newcommand {\bfalpha}{\bv{\alpha}}

\newcommand {\bfbeta}{\bv{\beta}}
\newcommand {\bfgamma}{\bv{\gamma}}
\newcommand {\bfzeta}{\bv{\zeta}}
\newcommand {\bfrho}{\bv{\rho}}

\newcommand{\reals}{\mathbb{R}}
\newcommand{\realsX}{\reals^{|\calX|}}
\newcommand{\realsXminusone}{\reals^{|\calX|-1}}
\newcommand{\realsK}{\mathbb{K}}
\newcommand{\realsKX}{\realsK^{|\calX|}}
\newcommand{\realsKplus}{\realsK_+}
\newcommand{\realsKplusX}{\realsKplus^{|\calX|}}

\newcommand{\calYSmall}{\calY_{\mathrm{small}}}
\newcommand{\calBK}{\calB_\realsK}

\newcommand{\colvec}[2]{\begin{bmatrix} #1 \\ #2 \end{bmatrix}}

\newcommand{\rcritical}{r_\mathrm{critical}}

\newcommand{\xmax}{x_{\max}}

\newcommand{\omegaup}{\overline\omega}
\newcommand{\omegadown}{\underline\omega}

\newcommand{\density}{\varphi}

\newcommand{\weightQTag}{M\pp{\calQ'(\bfalpha,r)}}
\newcommand{\weightVTag}{M\pp{\calV'}}

\newcommand{\bfy}{\bv{y}}

\newcommand{\bfybar}{\bv{\bar{y}}}
\newcommand{\bfz}{\bv{z}}

\newcommand{\calA}{\mathcal{A}}
\newcommand{\bfu}{\bv{u}}

\newcommand{\pd}{\mathop{}\!\partial} % For dx in integrals or derivatives

\newcommand{\calYTilde}{\tilde{\calY}}
\newcommand{\bfytilde}{\bfy^{R}}
\newcommand{\ytilde}{y^{R}}
\newcommand{\bfyutilde}{\bfy^{L}}
\newcommand{\yutilde}{y^{L}}

\newcommand{\dkl}{d_{\mathrm{KL}}}

\newcommand{\calYPunctured}{\calY_{\mathrm{punctured}}}

\newcommand{\conv}{\mathrm{conv}}

\newcommand{\yleft}{y_L}
\newcommand{\bfyleft}{\bfy_L}
\newcommand{\yright}{y_R}
\newcommand{\bfyright}{\bfy_R}